\documentclass[iop]{emulateapj}
\usepackage{graphicx}

\usepackage{amsmath}
\usepackage[]{natbib}
\usepackage[usenames]{color} 
\usepackage{ulem}
\newcommand{\kpc}{\>{\rm kpc}}

\newcommand{\kms}{\>{\rm km}\,{\rm s}^{-1}}
\newcommand{\kmsdex}{\>{\rm km}\,{\rm s}^{-1}\,{\rm dex}^{-1}}
\newcommand{\dexkpc}{\>{\rm dex}\,{\rm kpc}^{-1}}
\newcommand{\kpcdex}{\>{\rm kpc}\,{\rm dex}^{-1}}
\newcommand{\kmskpc}{\>{\rm km}\,{\rm s}^{-1}\,{\rm kpc}^{-1}}

\newcommand{\Msun}{\mbox{$\rm M_{\odot}$}}

\newcommand\degrees{^\circ}

\newcommand{\etal}{{et al.~}}
\newcommand\eg{{\it e.g.~}}
\newcommand\egc{{\it e.g.}}

\newcommand\ie{{\it i.e.~}}

\newcommand\vs{{vs.\ }}

\def\apj{{Astroph. J.}}
\def\apjs{{Astroph. J. Suppl.}}
\def\aj{{Astron. J.}}
\def\mnras{{MNRAS}}

\shorttitle{Genesis of the Milky Way's Thick Disk}

\shortauthors{Loebman \etal}
\begin{document}

\title{The Genesis of the Milky Way's Thick Disk via Stellar Migration}

\author{Sarah R. Loebman\altaffilmark{1}}
\author{Rok Ro\v{s}kar\altaffilmark{1}}
\author{Victor P. Debattista\altaffilmark{2,3}}
\author{\v{Z}eljko Ivezi\'{c}\altaffilmark{1,4}}
\author{Thomas R. Quinn\altaffilmark{1}}
\author{James Wadsley\altaffilmark{5}}

\altaffiltext{1}{Astronomy Department, University of Washington, Box 351580, Seattle, WA 98195-1580, USA; {\tt sloebman@astro.washington.edu}}
\altaffiltext{2}{Jeremiah Horrocks Institute, University of Central Lancashire, Preston, PR1 2HE, UK; {\tt vpdebattista@uclan.ac.uk}}
\altaffiltext{3}{RCUK Fellow}
\altaffiltext{4}{University of Zagreb, Croatia}
\altaffiltext{5}{Department of Physics and Astronomy, McMaster University, Hamilton, Ontario, L8S 4M1, Canada}

\begin{abstract} 

  The separation of the Milky Way disk into a thin and thick component
  is supported by differences in the spatial, kinematic and metallicity distributions
  of their stars. 
  These differences have lead to the predominant view that the thick disk
  formed early via a cataclysmic event and constitutes fossil evidence
  of the hierarchical growth of the Milky Way.  We show here, using
  $N$-body simulations, how a double-exponential vertical structure, with stellar
  populations displaying similar dichotomies can arise purely through
  internal evolution.  In this picture, stars migrate radially, while
  retaining nearly circular orbits, as described by Sellwood \& Binney
  (2002).  As stars move outwards their vertical motions carry them to
  larger heights above the mid-plane, populating a thickened
  component.  Such stars found at the present time in the solar
  neighborhood formed early in the disk's history at smaller radii
  where stars are more metal-poor and $\alpha$-enhanced, leading to
  exactly the properties observed for thick disk stars.  Classifying
  stars as members of the thin or thick disk by either velocity or
  metallicity leads to an apparent separation in the other property as
  observed.  This scenario is supported by the SDSS observation that
  stars in the transition region do not show any correlation between
  rotational velocity and metallicity. Although such a correlation is present in young
  stars because of epicyclic motions,  the radial migration
  mixes stars, washing out the correlation.  Using the Geneva
  Copenhagen Survey, we indeed find a velocity-metallicity correlation
  in the younger stars and none in the older stars.  We predict a
  similar result when separating stars by [$\alpha$/Fe]. The good
  qualitative agreement between our simulation and observations in the
  Milky Way are especially remarkable because the simulation was not
  tuned to reproduce the Milky Way, hinting that the thick disk may be
  a ubiquitous galaxy feature generated by stellar migration.  Nonetheless, we cannot exclude that
  some fraction of the thick disk is a fossil of a past more violent
  history, nor can this scenario explain thick disks in all galaxies,
  most strikingly those which counter-rotate with respect to the thin
  disk.

\end{abstract}


\section{Introduction}
\label{s:intro.tex}
In the years since \citet{Gilmore1983} first proposed a two component 
structure to the Milky Way disk, a large body of observational work 
has provided supporting evidence for contrasting thin and thick disk 
attributes.  
Structurally, the thin disk scale height is shorter than thick disk scale 
height 
\citep[for reviews see][and references therein]{Reid1993, Buser1999, Norris1999}, and the thick disk may have a longer scale length 
than the thin disk \citep{Robin1996, Ojha2001, Chen2001, Larsen2003}.
Kinematically, thick disk stars have larger velocity dispersions and lag the 
net rotation of the disk 
\citep{Nissen1995, Chiba2000, Gilmore2002, Soubiran2003, Parker2004, Wyse2006}.
Additionally, thick disk stars are older and metal poor relative to their thin 
disk counterparts \citep[\egc][]{Reid1993, Chiba2000, Bochanski2007a} and at a 
given iron abundance thick disk stars are $\alpha$-enhanced 
\citep{Fuhrmann1998, Prochaska2000, Tautvaisiene2001, Bensby2003, 
Feltzing2003, Mishenina2004, Brewer2004, Bensby2005}.
Moreoever, thin and thick disk attributes are not unique to the Milky Way 
but a ubiquitous feature for late type galaxies 
\citep{Burstein1979, vanderKruit1981, Abe1999, Neeser2002, Yoachim2005, Yoachim2006, Yoachim2007}. 

Recently, several SDSS--based studies have provided further strong
observational constraints on the structural, kinematic and chemical
properties of stars in the solar cylinder.  \citet[hereafter J08]{Juric2008}
used a photometric parallax method on SDSS data to estimate distances
to $\sim$48 million stars and studied their spatial distribution.
Because SDSS provides accurate photometry, which enables reasonably
robust distances (10--15\%, \citealt{Sesar2008}), as well as faint
magnitude limits ($r<22$) and a large sky coverage (6500 deg$^2$), J08
were able to robustly constrain the parameters of a model for the
global spatial distribution of stars in the Milky Way.  The J08 model
is qualitatively similar to previous work (\eg \citealt{Bahcall1980})
which identifies a clear change of slope in the counts of disk stars
as a function of distance from the Galactic plane; this change in
slope is usually interpreted as the transition from the thin to thick
disk \citep{Gilmore1983, Siegel2002}.

\citet[hereafter I08]{Ivezic2008} further extended this global
analysis of SDSS data by developing a photometric metallicity
estimator and by utilizing a large proper motion catalog based on SDSS
and Palomar Observatory Sky Survey data \citep{Munn2004}. I08 studied
the dependence of the metallicity, [Fe/H] and rotational velocity,
$V_{\phi}$, of disk stars on the distance from the Galactic plane and
detected gradients of both quantities over the distance ranging from
several hundred parsecs to several kiloparsecs.  Such gradients would be
expected in a thin/thick disk decomposition where the thick disk stars
are a separate population defined by a bulk rotational velocity lag and
a lower metallicity compared to those of the thin disk.  However, such a
model would also predict a correlation between the metallicity and the
velocity lag, which is strongly excluded ($\sim$$7\sigma$ level) by
the I08 analysis (see Figure 17, I08).  In this work we turn to a more
sophisticated Galactic description --- an $N$--body model --- to
characterize stars within the SDSS volume and solve this puzzle.

Over the past few decades, $N$--body simulations have been used to provide 
supporting evidence for three distinct theories of thick disk formation: 
violent relaxation \citep{JonesWyse1983}, substructure disruption 
\citep{Statler1988}, and heating by satellites \citep{Quinn1993}.
Several works have recently redressed these ideas.
\citet{Brook2004} and \citet{Bournaud2009} formed a thick disk in situ at high 
redshift during gas-rich mergers, where star formation is triggered by the 
rapid accretion of gas; this result is consistent with the thick disk forming 
through violent relaxation of the galactic potential.
In contrast, \citet{Kazantzidis2008},
\citet{Villalobos2008}, and \citet{Villalobos2010} 
investigated substructure disruption 
by using a cosmologically derived satellite accretion history 
to perturb a Milky Way-like disk; subhalo-disk encounters increased the scale 
height of this disk at all radii effectively forming a thick disk.
Finally, \citet{Abadi2003} showed that by 
tidally stripping/accreting satellites, the majority of the oldest stars 
in the thick disk could have formed externally rather than in situ. 

In this work, we study a new method of formation: radial migration.
Radial migration due to scattering from transient spirals 
was first described by \citet{Sellwood2002}. In 
this model energy and angular momentum changes occur from 
interactions with transient spiral arms, which move 
stars  at the corotation resonance inward or outward in radius while 
preserving their nearly-circular orbits.
\citet[][R08ab hereafter]{Roskar2008, Roskar2008a} studied this phenomenon in N-body + Smooth Particle Hydrodynamic (SPH) 
simulations of disk formation, 
and showed that migrations are possible on short timescales. 
They explored the implications of radial mixing for stellar populations for a
variety of stellar systems, including the solar neighborhood.
Here we extend their work by highlighting the 
vertical evolution that occurs as a result of migration.

We note that in this paper, we are not testing the validity of the 
other models of formation.  
However, recently, \citet{Sales2009} proposed using the eccentricity of 
orbits of stars 
in the thick disk to constrain the thick disk's formation mechanism;
they presented the eccentricity distributions that result from four N-body 
simulations: \citet{Abadi2003}, \citet{Villalobos2008}, R08b, and 
\citet{Brook2004}.
They found that the distributions that result from heating, radial migration
and mergers all had a strong peak at low eccentricity 
($\epsilon \sim 0.2 - 0.3$), while 
the distribution that results from accretion is centered at 
higher orbital eccentries ($<$$\epsilon$$> \sim 0.5$).
Building on this, \citet{Wilson2010} studied the eccentricity of orbits of 
stars in the thick disk observed in the Radial Velocity 
Experiment (RAVE) \citep{Steinmetz2006} and found these results to be 
inconsistent with expectations for the pure accretion simulation. 
\citet{Ruchti2010} also leveraged $\alpha$ measurements from RAVE to 
conclude that the $\alpha$ enhancement of the metal-poor thick disk implies 
that direct accretion of stars from dwarf galaxies did not play a major role 
in the formation of the thick disk. 
Using SDSS DR7, \citet{Dierickx2010} showed that the eccentricity of orbits of 
stars in the thick disk implies the thick disk is unlikely to be fully 
populated by radially migrated stars.  
We note that we cannot exclude that some fraction of the thick disk is a fossil
of a past more violent history, nor can this scenario explain thick disks in 
all galaxies.  
However, in what follows, we show that a large fraction of the stars in the 
thick disk could have formed in situ and arrived at their present location via 
radial migration.

The outline of this paper is as follows: in \S\ref{s:simul}, we
present two simulations, one with substantial migration and the other
with relatively little migration.  When we compare these two
simulations we can show that migration can build a thick disk as first
conceived by Gilmore \& Reid (1983): a component with a scale-height
larger than that of the thin disk.  In \S\ref{s:obs} we qualitatively
compare the Milky Way-like  simulation (with migration) with the SDSS observations to show that they
match each other sufficiently well to pursue further
comparison.  In \S\ref{s:solar} we present a detailed comparison
between the simulation and the local SDSS volume focusing on the
reason for the lack of correlation between $V_{\phi}$ and [Fe/H]; in
Appendix~\ref{s:spagna} we reconsider recent observational claims
concerning the lack of correlation between $V_{\phi}$ and [Fe/H]. 
In \S\ref{s:obs_decomp} we use the simulation as a proxy for the Milky
Way to show that classifying stars as members of the thin or thick disk by either 
velocity or metallicity leads to an apparent separation in the other property 
as observed.
In \S\ref{s:schoenrich} we compare our results to recent theoretical work
that used semi-analytics to investigate how the solar neighborhood could have 
been shaped by radial migration and chemical evolution effects.
In \S\ref{s:predict} we explore the correlation between [$\alpha$/Fe]
and age to show the diagnostic power of [$\alpha$/Fe] as a stand-in
for age.  
Finally, in \S\ref{s:concl} we summarize our results and conclusions.

\section{Numerical Simulations}
\label{s:simul}

We analyze the results of an $N$--body + SPH simulation designed to
mimic the quiescent formation and evolution of a Milky Way--mass
galaxy following the last major merger.  The system is initialized as
in \cite{Kaufmann2007} and R08ab and consists of a rotating,
pressure--supported gas halo embedded in an NFW \citep{Navarro1997}
dark matter halo.  This simulation was evolved for 10 Gyr using the
parallel $N$--body$+$SPH code, GASOLINE \citep{Wadsley2004}.  As the
simulation proceeds, the gas cools and collapses to the center of the
halo, forming a thin disk from the inside--out.  
Gas is continually infalling from the hot halo onto the disk for the
duration of the simulation.
Star formation and
stellar feedback are modeled with subgrid recipes as described in
\citet{Stinson2006}.  Importantly, the stellar feedback prescriptions
include SN II, SN Ia and AGB metal production, as well as injection of
supernova energy which impacts the thermodynamic properties of the
disk interstellar medium (ISM).  Metal diffusion is calculated from a
subgrid model of eddy turbulence based on the local smoothing length
and velocity gradients \citep{Smagorinsky1963, Wadsley2008}.  The
simulation we utilize is nearly identical to R08ab (see R08ab for
further details), but with the addition of metal diffusion
\citep{Shen2009}.

No {\it a priori} assumptions about the disk's structure are made ---
its growth and the subsequent evolution of its stellar populations are 
completely spontaneous and governed only by hydrodynamics/stellar feedback 
and gravity. 
Although we do not account for the full cosmological context, merging
in the $\Lambda$CDM paradigm is a higher order effect at the epochs in
question \citep{Brook2005}. 
Thus, our model galaxy lacks some structural components such as a stellar 
halo, 
which in $\Lambda$CDM is built up primarily during the merging process 
\citep[\eg][]{Bullock2005, Zolotov2009}. 
Our focus here, however, is disk evolution; by simplifying our assumptions, 
we are able to use much higher resolution and more easily study the impact of 
key dynamical effects on observational properties of stellar populations 
within the disk.

\begin{figure}
\epsscale{1.15}
\vskip .1 in
\plotone{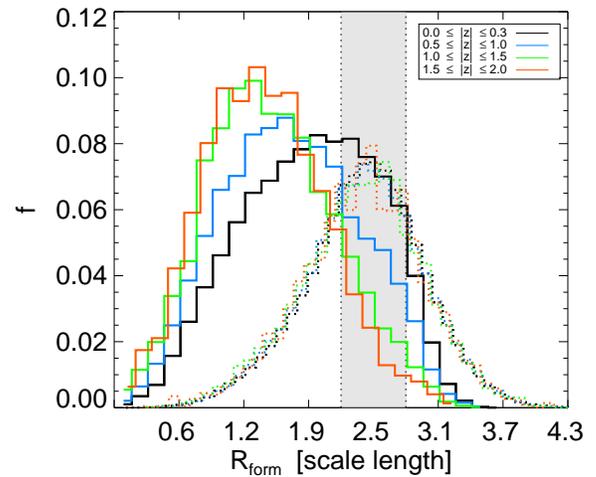}
\vskip .1 in
\caption{Star particles that fall within the solar cylinder at the end of 
the simulation are considered here.
These stars are broken into four volumes by 
distance away from the mid-plane, $|z|$, with low to medium to high given 
by black to blue to red. 
For comparison, four similar volumes from the LSB simulation with little
radial migration are over-plotted (dotted lines).
For each volume, the formation radius of stars is shown; in the MW simulation,
away from the mid-plane, a large fraction of the stars formed 
significantly interior to their final location. 
For reference, the solar cylinder is indicated in the shaded gray region: 
galactocentric radius $=2.2$--$2.8$ scale lengths.  
In the MW simulation this corresponds to
7 kpc $\leq R \leq$ 9 kpc while in the LSB simulation it corresponds 
to 11 kpc $\leq R \leq$ 14 kpc.}
\label{f:rform}
\end{figure}

\begin{figure}
\epsscale{1.15}
\vskip .1 in
\plotone{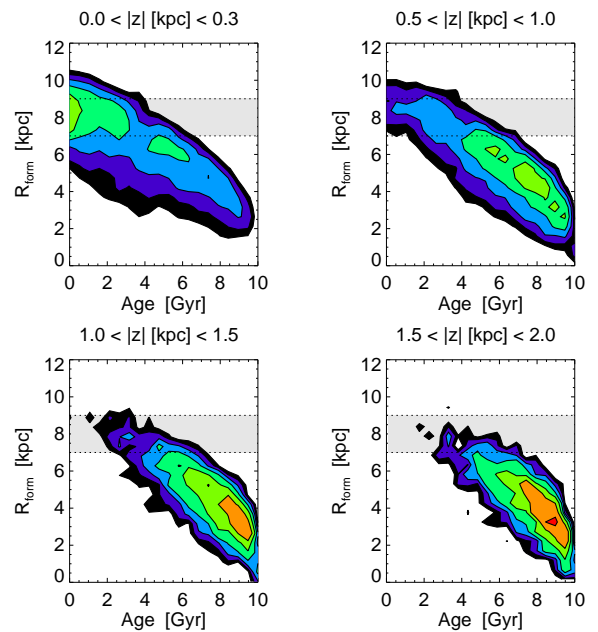}
\vskip .1 in
\caption{Contour plots of the MW simulation showing the distribution 
of $R_{form}$ $\vs$ $Age$
for the four volumes considered in Figure~\ref{f:rform} with 
solar cylinder shaded in gray.  
For all $z$, older stars formed significantly interior to their final location;
this net outward movement of stars over time is due to radial migration.  
Volumes sampling the thick disk ($|z| \ge 1 \kpc$) are dominated by older 
stars that have migrated to the solar radius from interior radii.}
\label{f:rform_cont}
\end{figure}

\begin{figure}
\epsscale{1.15}
\vskip .1 in
\plotone{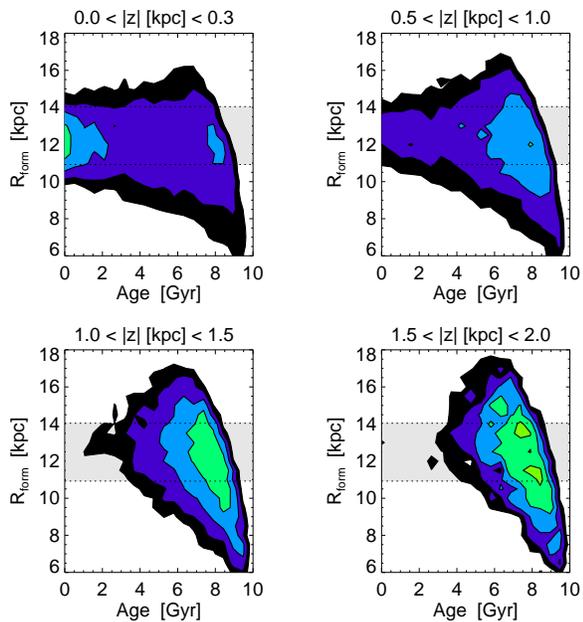}
\vskip .1 in
\caption{Same as Figure~\ref{f:rform_cont} but for 
the LSB simulation which has little radial migration.
Regardless of distance away from the midplane, 
all stars originate from a roughly symmetric distribution centered at 
the midpoint of the cylindrical volume.
While volumes sampling the thick disk ($|z| \ge 1 \kpc$) are dominated by 
older stars, these stars are largely uninfluenced by radial migration.}
\label{f:rform_cont_comp}
\end{figure}

Based on such simulations, R08ab presented the implications of 
stellar radial migration resulting from the interactions of stars with 
transient spiral arms \citep{Sellwood2002} on the observable properties of 
disk stellar populations. 
Radial migration efficiently mixes stars throughout the disk into the solar 
neighborhood, resulting in a flattened age-metallicity relation (R08ab).
Figures~\ref{f:rform},~\ref{f:rform_cont}, \& ~\ref{f:rform_cont_comp} 
illustrate the basic premise of this paper --- stars migrate radially and 
in the process rise out of the plane over time, 
so many stars are presently not near their birth place.  
Previous studies have shown \citep{Loebman2008, Sales2009, Caruana2009, Schoenrich2009} that
the vertical evolution that results from radial migration can influence the 
characterization of the thick disk. 

In order to further illustrate the importance of radial migration
within our adopted Milky Way (MW) simulation, we
have repeated much of our analysis on a control case. 
The control simulation is a system with the 
same initial conditions as the MW simulation except for having a higher 
angular momentum content with a dimensionless spin parameter $\lambda = 0.1$ 
\citep{Bullock2001}.  
This results in a more extended disk 
(final disk scale-length $= 5.04$ $\kpc$, versus $3.23$ $\kpc$ 
for the MW simulation), 
possibly similar to a low surface brightness galaxy; 
we therefore refer to this simulation as the LSB simulation.  
Due to its lower surface density, the disk forms weaker spirals and 
as a result the stellar populations at all radii are less affected by 
radial mixing.  
When we compare migration as a function of scale-lengths, 
we find that there is significantly less migration in the LSB galaxy 
than in the MW galaxy. 

\begin{figure}
\epsscale{1.2}
\plotone{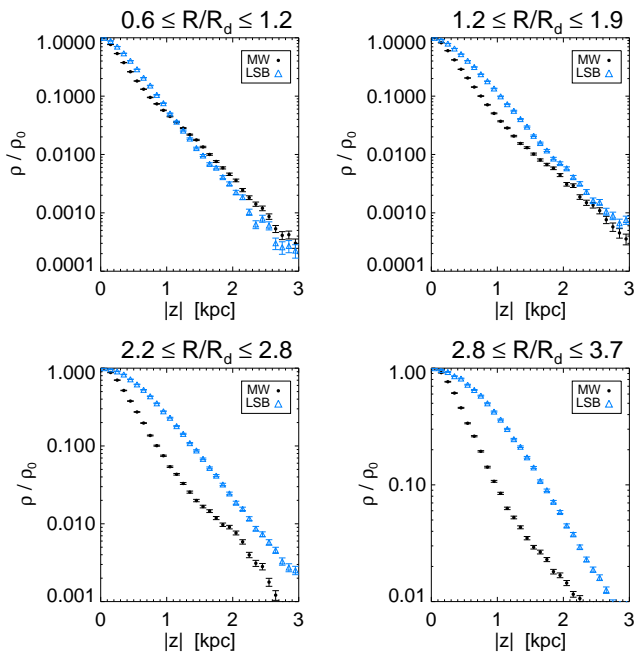}
\caption{Density profiles drawn from analogous regions within the 
         MW and LSB simulations. At larger radii, the LSB simulation has a 
         fairly flat (pure exponential) profile whereas the MW simulation has 
         a transition between a steep and shallow profile.}
\label{f:density_comp}
\end{figure}

The distribution of stellar mass away from the midplane is strongly
affected by radial migration; this can be seen in
Figure~\ref{f:density_comp}, which contrasts the MW simulation against
the LSB case.  Here the normalized mass density distribution within
four analogous cylindrical volumes drawn from a variety of radii are
presented.  At larger radii, the steepness of the profiles are quite
different; the LSB simulation has a constant slope while the MW
simulation shows a transition from a steep to a shallow density
distribution.  Thus the MW simulation cannot be characterized by a
single exponential or sech$^2$ component in the vertical direction, as
we show explicity in the following Section and
Figure~\ref{f:density_profiles}.  
It is this double-component nature
which first led to the identification of the thick disk \citep{Gilmore1983}; we have thus
shown that this feature need represent nothing more than internal
evolution of the Milky Way.

\section{Comparison of Simulations with SDSS}
\label{s:obs}

In the following section, we compare SDSS observations with the MW
simulation to demonstrate its usefulness as a model for understanding
the Milky Way thick disk.  Here we study the stellar mass
distribution, rotational velocity and
metallicity as functions of distance from the Galactic
plane, $|z|$, and galactocentric cylindrical radius, $R$.  We draw
qualitative comparisons between the datasets by examining their mass weighted
metallicity and kinematic distributions in this $R$--$|z|$ space.

\begin{figure}
\epsscale{1.2}
\plotone{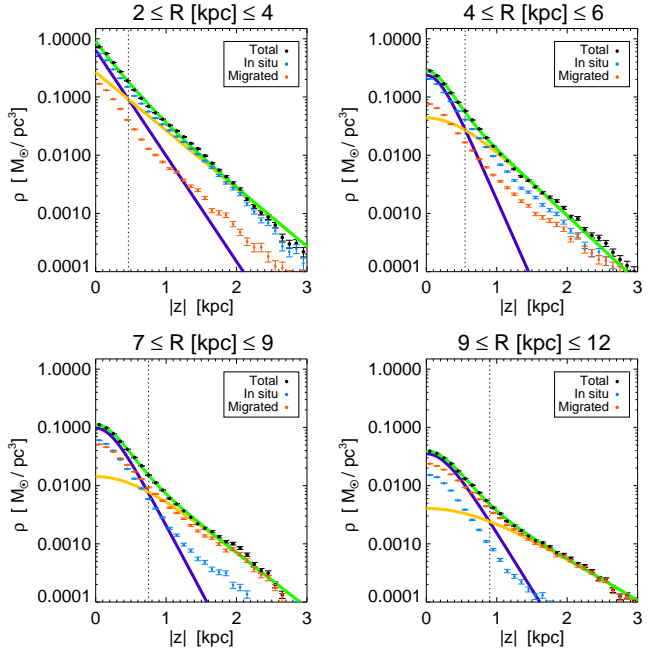}
\caption{Simultaneous fits to the vertical density profile in the MW simulation for radial bins: 
         $R = 2 - 4$ $\kpc$, $R = 4 - 6$ $\kpc$, 
         $R = 7 - 9$ $\kpc$, $R = 9 - 12$ $\kpc$.
         While the radial bin sampling the smallest radii is best fit 
         by a double exponential function, all other bins are better fit 
         by a double sech$^2$ function. 
         Best fit parameters are given in Table~\ref{t:best_fit}. 
         Thin, thick, and total curves shown in purple, orange and 
         green repectively; the vertical dotted line marks 
         the intersection between thin and thick disk components.
	 Red and cyan points represent stellar mass density that has migrated 
         more than $2 \kpc$ or less than $2 \kpc$ respectively from radius of 
         formation.  
         }
\label{f:density_profiles}
\end{figure}

\begin{table}[!h]
\begin{center}
\begin{tabular}{lcccr}
\hline
R [$\kpc$]&N1 [\Msun/$pc^3$]&h1 [pc]&N2 [\Msun/$pc^3$]&h2 [pc]\\
\hline
$2 - 4$&0.638&239&1.000&266\\
$4 - 6$&0.237&316&0.044&763\\
$7 - 9$&0.098&381&0.014&913\\
$9 - 12$&0.035&444&0.004&1197\\ \hline
\end{tabular} 
\end{center}
\caption{Best fit parameters to radial bins sampled in 
         Figure~\ref{f:density_profiles} given by: 
         thin disk normalization ($N1$), thin disk scale height ($h1$), 
         thick disk normalization ($N2$), \& thick disk scale height ($h2$).}
\label{t:best_fit}
\end{table}

\begin{figure*}[!th]
\epsscale{1.1}\plotone{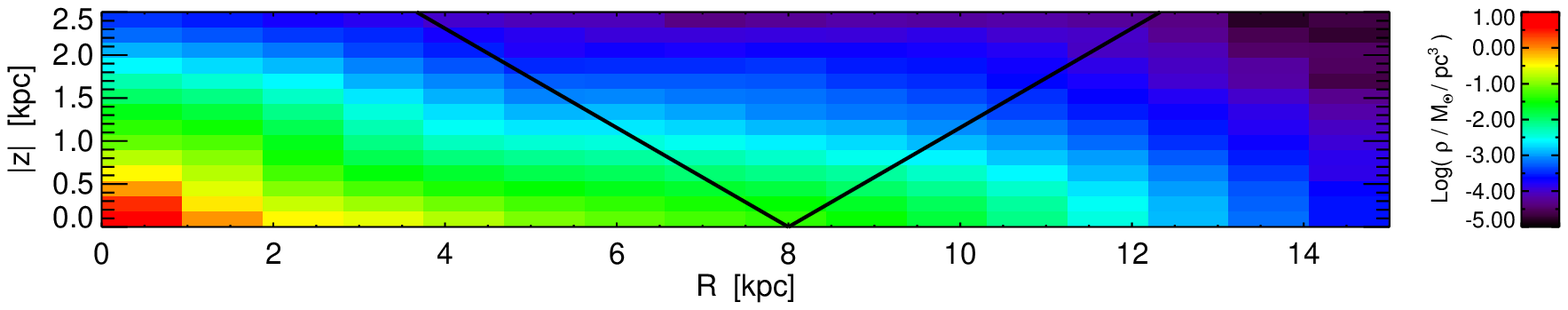}
\vspace*{.3cm}
\epsscale{1.1}\plotone{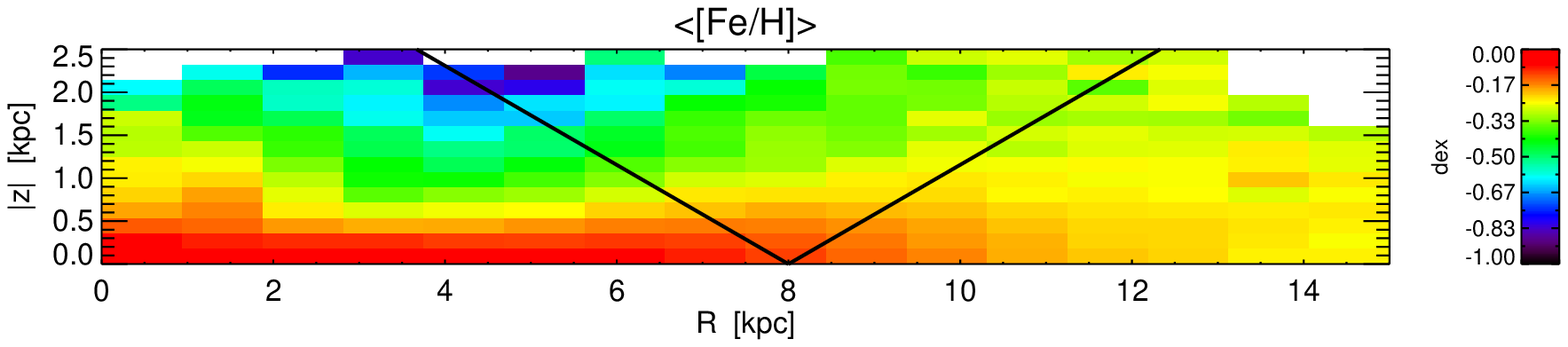}
\vspace*{.3cm}
\epsscale{1.1}\plotone{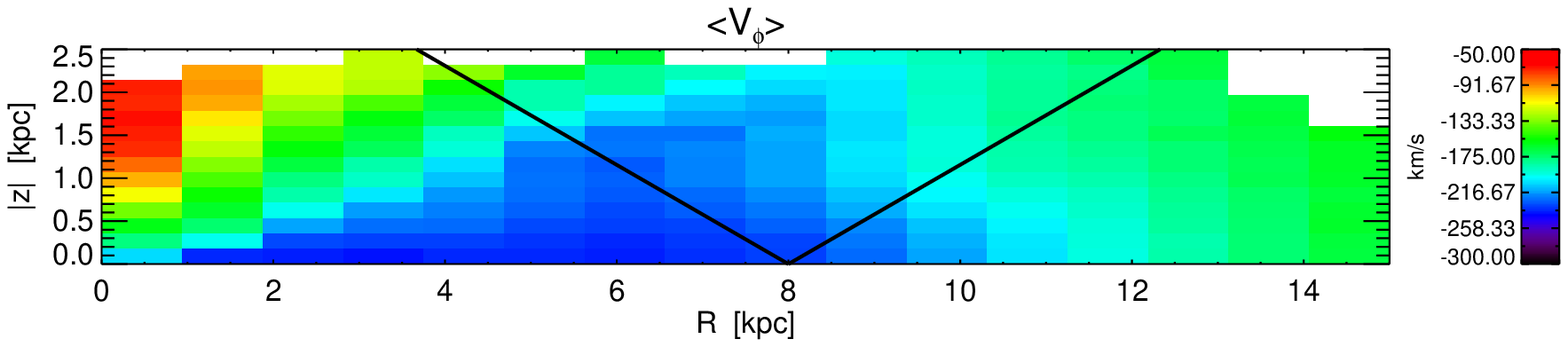}
\caption{ 
  Top:    Mass density distribution in galactocentric coordinates of
          the entire MW simulation.
          The colors scale logrithmically with density in each bin.
          Overplotted is the SDSS field of view ($|b| > 30 \degrees$).
  Middle \& Bottom: Mass-weighted mean [Fe/H] and $V_{\phi}$ values as mapped 
          onto the $R$--$|z|$ plane.
          For a color box to be plotted we required a minimum of 50 star particles.
}
\label{f:color_maps}
\end{figure*}

The observed Milky Way disk is best fit by a 2-component model
that is exponential both in the $R$ and $z$ directions (see Table 10, 
bias-corrected results, J08).
The top panel of Figure~\ref{f:color_maps} shows the 
mass weighted density distribution of the entire MW simulation at its 
final timestep.  This distribution is in qualitative agreement 
in both the $R$ and $z$ directions with J08 for up to $\sim$$2.5$ $\kpc$ above 
the disk's plane and $\sim$$15$ $\kpc$ from the galactic center. 

Figure~\ref{f:density_profiles} shows our best fits to the vertical
density profiles for the MW simulation for radial bins $R = 2 - 4$
$\kpc$, $R = 4 - 6$ $\kpc$, $R = 7 - 9$ $\kpc$ and $R = 9 - 12$
$\kpc$.  As in J08, the innermost radial bin is best fit by a double
exponential function; however, all other bins are better fit by the
sum of two double sech$^2$ profiles, which is in agreement with the
theoretical work by \citet{Spitzer1942} and observational results for other
galaxies by
\citet{Yoachim2006}.  Despite the small discrepancy between sech$^2$
\vs exponential fits, the simulation is in good qualitative agreement
with the SDSS-based analysis of the Milky Way (J08).

In addition, Figure~\ref{f:density_profiles} shows the vertical
density profiles as divided into two populations: 
radial migrators ($|R - R_{form}| > 2 \kpc$) and 
an in situ population ($|R - R_{form}| \leq 2 \kpc$).
We note the second component fit for radial bins 
$R = 7 - 9$ $\kpc$ and $R = 9 - 12$ $\kpc$ is entirely 
dominated by star particles that have migrated.
We expect that stars that moved from elsewhere combined with stars that were
born locally should not naturally conspire to produce a single continuous
profile.

For the solar cylinder, $R = 7 - 9 \kpc$, we find that the model
distribution of stars as a function of $|z|$ resembles a double
sech$^2$ profile, with the ``transition'' height of $|z|\sim$$0.75
\kpc$, comparable to $\sim$$1 \kpc$ found by J08 for the Galactic
disk.  We have found the best-fit scale heights to be 381 pc and 913
pc, in qualitative agreement with the best-fit scale heights of 270 pc
and 1200 pc for the SDSS data (J08).  The scale height ratio suggested
by the simulation is slightly low --- $\sim$$2.4$, instead of $\sim$$3$
from the data.  Moreover, the simulation's thick disk to thin disk
normalization is slightly discrepant --- $\sim$$0.14$, rather
than $\sim$$0.12$ suggested by the SDSS data.  Overall, we find that
the thick disk in the MW simulation, formed through the process of
radial migration, is qualitatively very similar to the observed
Galactic thick disk.  As Figure \ref{f:density_comp} clearly attests,
this conclusion does not apply to the LSB simulation. 

The median metallicity of the Milky Way disk exhibits a clear vertical
gradient (see Figure 9, bottom two panels, I08).   
Notably, the spatial variation of the median metallicity does not follow the
distribution of the stellar number density (I08). 
The middle panel of Figure~\ref{f:color_maps} shows that 
the MW simulation reproduces a qualitatively similar 
metallicity distribution; we note that a constant additive offset of 0.2 dex has been 
applied to [Fe/H] throughout the simulation so that the median 
metallicity in the plane of the disk in the solar cylinder matches 
observations. 
As expected, at low galactic latitudes and small radii, the
volume is dominated by high (near solar) metallicities.  At higher
latitudes the volume is increasingly metal poor.

As with metallicity, previous studies have found a gradient 
in the median $V_{\phi}$ with respect to $z$ in the Milky Way
(see Figures 5, 8, and 9 in Bond \etal
(2010, hereafter B10)\nocite{Bond2010}, and references therein).
These authors concluded that $V_{\phi}$ is also well characterized by a 
non-Gaussian distribution 
\citep[see][for the implied distribution function]{Binney2010}.
The bottom panel of Figure~\ref{f:color_maps} shows that the MW 
simulation reproduces qualitatively similar $V_{\phi}$ properties, 
including a strong $|z|$ gradient.  

Thus on a gross scale the MW simulation qualitatively matches patterns 
observed in mass density, metallicity and rotational velocity in 
the Milky Way disk.  
We now look at the solar cylinder in greater detail.

\section{Effects of Radial Migration on the Solar Cylinder}
\label{s:solar}

\subsection{Vertical Gradients}
\label{s:grad}

Here we study in detail the distributions of age, stellar mass,
rotational velocity and metallicity as functions of $|z|$ for the
solar cylinder.  To be consistent with the analysis of high galactic
latitude SDSS data by J08 and I08, we select model particles from an
annulus with 7 $\kpc$ $\leq R \leq$ 9 $\kpc$.  This radial cut spans
2.2--2.8 scale lengths from the center of model galaxy (scale length
$3.2$ $\kpc$), and covers the Sun's location of $\sim$3 disk scale
lengths from the Milky Way center (scale length $\sim$$2.6 \kpc$) (see
bias-corrected value, Table 10, J08).

\begin{figure}[!h]
  \epsscale{.75}
  \plotone{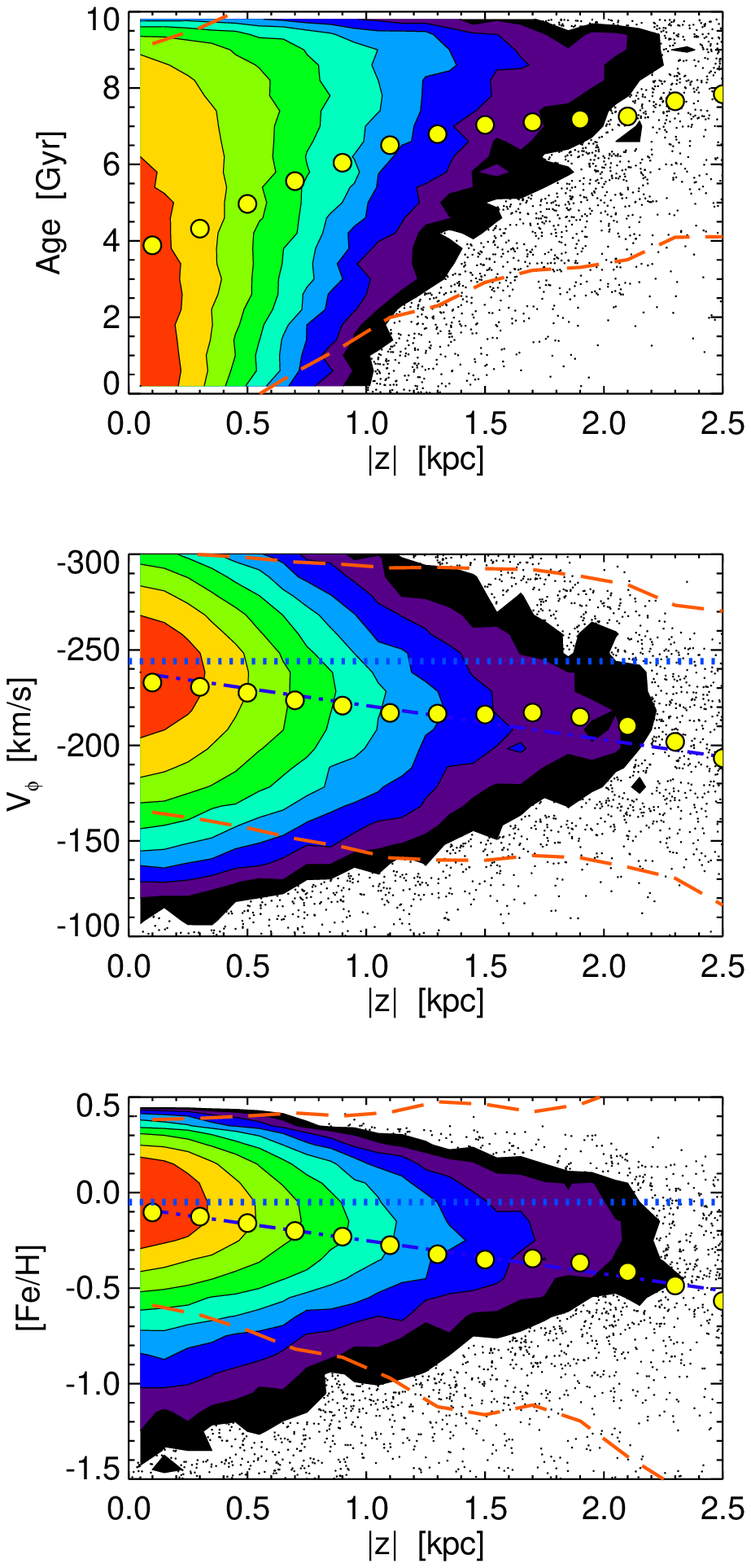}
  \caption{The behavior of $\sim$ $200,000$ model particles selected from a 
           galactocentric cylindrical annulus with 7 $\kpc$ $<R<$ 9 $\kpc$. 
	   Age, rotational velocity, and metallicity are shown as a 
	   function of height in the top, middle, and bottom panels 
           respectively.
	   The data is represented by color-coded contours (low to medium to 
           high: black to green to red) in the regions of high density,
	   and as individual points otherwise. 
	   The large symbols show the weighted mean
	   values in $|z|$ bins, and the dashed lines show a 2$\sigma$ 
	   envelope around the weighted means. 
	   The dot-dashed line shows the best linear fit to these means. 
	   Overplotted for reference in a dotted blue line is the mean 
           rotational velocity and mean metallicity at $z=0$ 
           in the MW simulation.}
\label{f:ringberg}
\end{figure}

The behavior of the MW simulation in this particular volume 
is illustrated in Figure~\ref{f:ringberg}. 
The top panel shows the age distribution as a function of $|z|$.
As expected, near the midplane, the population is dominated by young
stars, and
the mean age monotonically increases with increasing distance  
from the midplane.
Only very old stars are found at large $|z|$: by $|z| \sim$$0.5 \kpc$ 
the mean age is already $\sim$5 Gyr.
Recall that Figure~\ref{f:rform_cont} shows that with increasing distance 
from the midplane, 
the stellar population becomes dominated by older stars that 
formed closer to the center of the disk.
These two figures taken together give a coherent picture of the net dynamic 
effect on the system: 
{\it on average stars move both radially outward and away from the midplane over 
time.} 

Radial migration is able to change the extent of a star's vertical oscillation,
as a well as its rotational velocity (middle panel,
Figure~\ref{f:ringberg}).  In the MW simulation we find a
vertical gradient of rotational velocity of -17 $\kmskpc$ 
(compared to a gradient of -30 $\kmskpc$ found by I08).  Note that
stars which significantly lag the rotation of the disk in the midplane
have traditionally been regarded as members of the thick disk.

Another observed trend that is physically well motivated by the net outward 
and upward movement of stars over time is the decline of metallicity with 
increased height (bottom panel, Figure \ref{f:ringberg}). 
In the MW simulation, the metallicity distribution changes with $|z|$ with a best-fit gradient of $\sim$$0.19$ $\dexkpc$, again in qualitative agreement with the measured value for the Milky Way of $\sim$$0.30$ $\dexkpc$ (I08).

To understand how this trend arises, recall that the top panel of 
Figure \ref{f:ringberg} shows that the stellar population away from the plane
is dominated by old stars.
As Figure \ref{f:rform_cont} shows, 
the oldest stars at large $|z|$ mostly formed in the inner $2 - 4$ $\kpc$.
Moreover, at early times, the radial metallicity gradient was steep 
(see Figure 2 in R08b); stars that are now in the thick disk in the solar 
cylinder in the MW simulation were once at the outer edge of the forming 
thin disk and hence formed at a low metallicity.
As a result, stars that formed at a radius of $2 - 4$ $\kpc$ early on in the 
galactic history are necessarily metal poor. 
These stars have been subsequently moved out and up over time.
At later times, the metallicity gradient flattened out as the disk grew. 
Again, looking at Figure \ref{f:rform} (top right panel), 
one can see a significant fraction of young stars (\ie less than 4 Gyr old) 
formed at R $\sim$6 $\kpc$ where there was a relatively more metal rich ISM. 
These stars had less time to migrate and as a result remain closer to 
the midplane of the disk. 
This complex co-dependence of radial migration, birth location,
and metallicity gradient evolution then gives rise to a vertical 
metallicity gradient in the solar neighborhood.

\subsection{Thin/Thick Disk Transition}

We turn our attention now to a particular region within the solar cylinder: 
$|z| = 0.5 - 1.0$ $\kpc$, $R = 7 - 9$ $\kpc$.  
This region in the simulation is analogous to the  
thin--thick disk ``transition zone'' considered by I08: 
within this volume a roughly equal number of thin and thick disk stars 
is expected (see bottom left panel, Figure \ref{f:density_profiles}). 
The transition region within the MW simulation occurs in approximately the same 
place as in the data (0.75 $\kpc$ versus $\sim1$ $\kpc$)  
which allows us to draw a direct comparison to the slice analyzed in I08.
We show here that all the trends observed in the SDSS data can be explained 
by a continuous distribution rather than two distinct populations.

In Figure \ref{f:age_plots} we show the distribution of observable properties 
as a function of age within this slice of the simulation. 
The top left panel of Figure \ref{f:age_plots} shows the distribution of ages 
in this thin volume slice: the region is well populated ($\sim$30,000 
star particles) and predominantly old ($36\%$ older than 7 Gyr, $63\%$ 
older than 5 Gyr). 
The remaining panels in  Figure \ref{f:age_plots} illustrate probability 
densities of formation radius, metallicity, and rotational 
velocity versus age.
In all cases, the distributions do not suggest distinct populations.
For the given $|z|$ slice, stars older than about 4 Gyr are both numerous 
and formed significantly interior to their present location (top right panel). 
These stars also show a rotational velocity lag of $\sim$$20 \kms$, as shown
in the bottom left panel. The oldest stars, those with ages $\ga 8$ Gyr, 
have significantly lower metallicities than younger stars (bottom right panel).
And as noted previously, the oldest stars formed at a range of interior radii, when the ISM 
metallicity was low and had a steep radial gradient.
These effects drive the behavior of various observables, as discussed next.

\subsection{Rotational Velocity \vs Metallicity}

We have shown in section \S\ref{s:grad} that the MW simulation yields vertical 
trends in metallicity and rotational velocity similar to those found in SDSS 
observations presented in I08 \& B10. 
SDSS data also revealed a surprising lack of correlation between $V_{\phi}$ 
and [Fe/H], contrary to the expectations based on a traditional 
two-disk model (see Appendix~\ref{s:spagna} for a discussion of competing observational claims). 
In the top left panel of Figure \ref{f:vphi_feh_decomp}, we show 
$V_{\phi}$ \vs [Fe/H] for the vertical slice considered 
above ($|z| = 0.5 - 1.0 \kpc$) corresponding to the transition between the 
thick and thin disk in the simulation. 
The MW simulation also yields a lack of correlation between these 
quantities: although both rotational velocity and metallicity show vertical 
gradients, when stars are selected from a thin $z$ slice, velocity and 
metallicity are not correlated.

\begin{figure}[!h]
\epsscale{1}
    \plotone{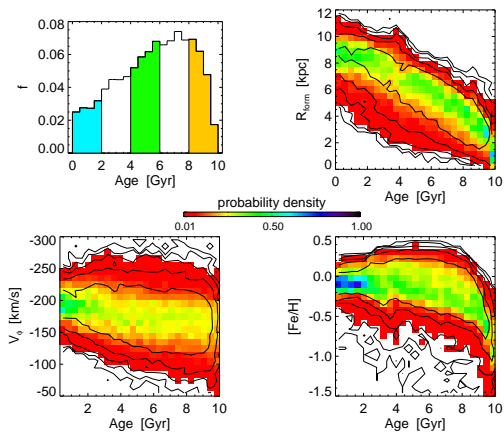}
    \caption{Top left: histogram of stellar ages for subset of data from 
             within thin slice spanning 
             $|z| = 0.5 - 1.0$ $\kpc$, $R = 7 - 9$ $\kpc$; shaded regions
             correspond to age slices considered in 
             Figure~\ref{f:vphi_feh_decomp}.
             The remaining three panels illustrate probability density maps 
             with logarithmically spaced contours overplotted; here each 
             column sums to 1. 
             Clockwise from top right: formation radius, metallicity, and 
             rotational velocity as a function of age.  
             Note, the top right panel is the same considered in the top right 
             panel in Figure~\ref{f:rform}.}
    \label{f:age_plots}
\end{figure}

We can understand why there is no correlation between $V_{\phi}$ and
[Fe/H] if we consider the thin slice as occupied by an age ensemble
that was brought together by radial migration.  In this light, we can
decompose the sample by age to see how the behavior of each population
is modified with time (see top left panel of Figure~\ref{f:age_plots}
for the age sub-samples considered).  

The top right panel of Figure
\ref{f:vphi_feh_decomp} shows a strong correlation
between rotational velocity and metallicity exists for young stars;
the gradient is $-29$ $\kmsdex$.  The young stars constitute a small
fraction of the overall population in the selected volume ($10\%$)
and, as the top right and bottom right panels of Figure
\ref{f:age_plots} illustrate, young stars have correlated metallicity
and formation radius.  Thus if a young star has a metallicity
different than the surrounding ISM, it must originate from somewhere
outside the solar cylinder and be at or near perigalacticon (if
$R_{g}$ $\geq$ $9$ $\kpc$, where $R_g$ is the radius of the stellar
guiding center) or at or near apogalacticon (if $R_{g}$ $\leq$ $7$
$\kpc$); consequently, it will either lead or lag the local
standard of rest (LSR).

For increasingly older age bins, the gradient diminishes and
ultimately fades away.  The bottom left and right panels of Figure
\ref{f:vphi_feh_decomp} show $V_{\phi}$ \vs [Fe/H] for intermediate (4
Gyr $\leq$ Age $\leq$ 6 Gyr) and old (8 Gyr $\leq$ Age $\leq$ 10 Gyr)
populations respectively.  The intermediate age stars constitute
$24\%$ of the stellar population in the thin slice and retain a slight
gradient $ = -17$ $\kmsdex$.  The oldest stars make up $21\%$ of the
stellar population and show a nearly flat/slightly positive slope of
$8$ $\kmsdex$.

The peak in the distribution of intermediate and oldest age star
particles is offset to progressively lower metallicity and rotational
velocity.  The metallicity of intermediate age stars ranges from $-0.5
<$ [Fe/H] $< 0.5$ while the metallicity of old stars range from $-1.0 <$
[Fe/H] $< 0.0$.  We can think of these three age bins as dominating
different portions of the overall $V_{\phi}$ \vs [Fe/H]
space, with the space spanned by intermediate and old stars being
perpendicular to the peak in the correlation in young stars. 
Thus,  for the data cut into slices of [Fe/H], 
 the distributions of $V_{\phi}$ are offset relative to one another.

For the highest density contours, the mean value of the oldest stars is 
[Fe/H]$= -0.5$, $V_{\phi}=-210 \kms$; this is significantly lower than the 
mean value in the highest density contour of the full age sample 
[Fe/H]$= -0.1$, $V_{\phi}=-225 \kms$.
The full age sample's peak mean value matches the intermediate age values, 
which is in turn lower than the peak of the youngest stars 
[Fe/H]$= -0.05$, $V_{\phi}=-245 \kms$.
When the entire population is considered as a whole, the correlations 
disappears.

Why does the $V_{\phi}$-[Fe/H] gradient diminish with increasing
age?  To understand this, we return to Figure \ref{f:age_plots} and
recall that most of the stars within this thin cut are intermediate to
old age, and they did not form within $7$ $\kpc$ $\leq$ $R$ $\leq$
$9$ $\kpc$.  A range of formation radii corresponds to a range of
formation environments; this maps to a range of resulting
metallicities for stars of a given old age bin.  At the same time,
depending on a star's formation location and subsequent migration and
scattering off the disk substructure, it can end up with a range of
possible rotational velocities (see bottom left panel, Figure
\ref{f:age_plots}).  Hence, even a single old age bin samples a wide
range of formation environments and rotational velocities resulting
from unique dynamical histories, which are not directly correlated.

We stress the significant difference between these results and the
traditional double-disk interpretation: in the latter case a correlation 
between $V_{\phi}$ and [Fe/H] is expected as one moves from 
younger, metal rich thin disk population to an older, metal poor,
thick disk population. In contrast, here we find that a trend is present in
the young stars but absent in the older stars.  The trend in
the young stars can be easily understood as arising from epicyclic
motions of stars with their birth radii imprinted into their
metallicities.  As migration moves the guiding centers of stars, this
metallicity encoding is erased, and the correlation between $V_{\phi}$
and [Fe/H] disappears.

\subsection{Geneva Copenhagen Survey}

\begin{figure}[!h]
  \epsscale{1}
  \plotone{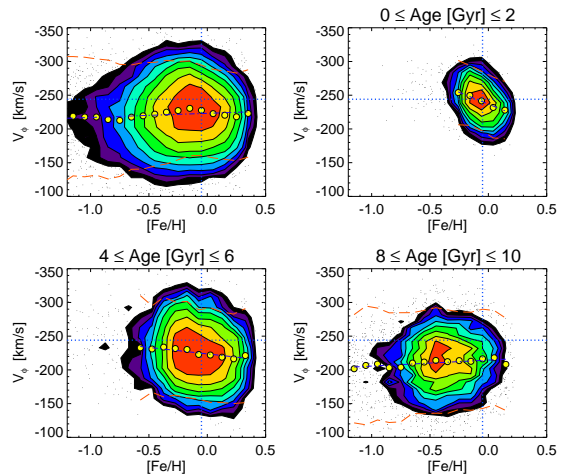}
  \caption{Decomposition of rotational velocity \vs metallicity by age. 
           Top left figure, rotational velocity \vs metallicity for all 
           stars within thin slice 
           $|z| = 0.5 - 1.0$ $\kpc$, $R = 7 - 9$ $\kpc$. 
           Rotational velocity and metallicity are not correlated.
           Top right, bottom left, and bottom right panels show stars with 
           ages 1, 5, 9 Gyr $\pm 1$ Gyr respectively.
           Overplotted for reference with dotted blue lines is the mean 
           rotational velocity and mean metallicity of the gas 
           within the MW simulation's solar cylinder. 
           As the top left panel of Figure~\ref{f:age_plots} illustrates, 
           the youngest stars are a small fraction of the overall mass 
           distribution; however, these stars show a clear trend of lower 
           metallicity at higher rotational speed.
           This trend diminishes and eventually disappears for increasingly 
           older stars.
           Additionally, the mean trend for each age slice is slightly offset;
           upon superposition, all evidence of any pre-exisiting trend 
           is erased.}
  \label{f:vphi_feh_decomp}
\end{figure}

As we demonstrated above, we can understand why there is no correlation
between $V_{\phi}$ and [Fe/H] when we decompose the sample by age.
Older populations have had more time to radially mix; thus in the
older age bins the expected trend disappears. That is, the evolution of 
the trends shown in Figure~\ref{f:vphi_feh_decomp} is a unique signature
of radial mixing taking place in the disk. 

We can verify whether this is the case in the solar neighborhood by
utilizing observational data taken from the Geneva-Copenhagen Survey
(GCS) \citep{Holmberg2009}. 
The GCS samples a wide range of stellar ages and is reasonably well populated
with old stars to make qualitative comparsions to the MW simulation in 
the plane of the disk.
While what we have shown in Figure~\ref{f:vphi_feh_decomp} is the
relationship at 0.5-1 $\kpc$ above the plane, these trends evolve in
the same way as a function of age in the mid-plane.  

\begin{figure}[!h]
  \epsscale{1}
  \plotone{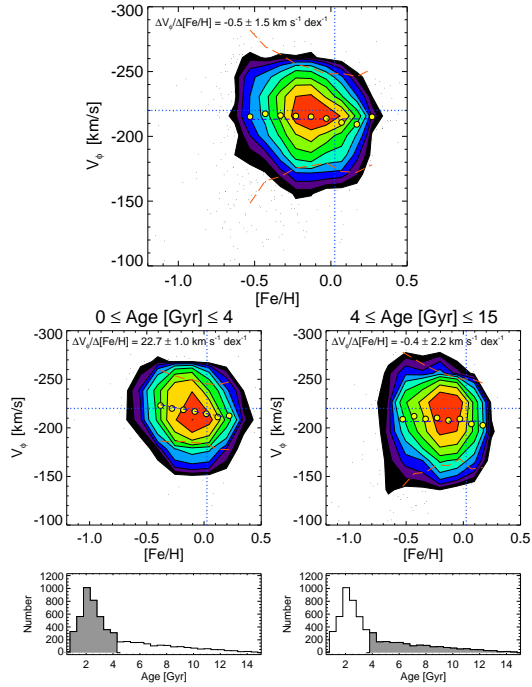} 
  \caption{$V_{\phi}$ \vs [Fe/H] for stars from the Geneva Copenhagen 
           Survey (GCS), which samples stars within 100 pc of the Sun. 
           We have repeated the analysis we presented in 
           Figure~\ref{f:vphi_feh_decomp} for all GCS stars flagged as 
           non-binary.
           Top panel: when the sample is considered as a whole, 
           there is no discernable trend.
           Middle row: the sample split into two broad age bins:
           Age = 0-4 Gyr and Age = 4-15 Gyr.  
           Left middle panel, the young stars show a trend, 
           while in the right middle panel, the old stars are not correlated.  
           This result agrees with the prediction from the MW simulation 
           and strongly motivates a history of radial mixing in the 
           solar neighborhood. 
           Bottom panels: histogram of overall age distribution, 
                          with shaded region corresponding to the data 
                          sampled in the panel above it.}
  \label{f:gc}
\end{figure}

We have selected the non-binary stars from the GCS and repeated the
analysis we presented above.  Figure~\ref{f:gc} shows the results: the
top panel shows that there is no net trend when the sample is
considered as a whole.  Splitting the sample into two broad age bins (0-4
Gyr and 4-15 Gyr) yields a trend only in the young stars -- the
correlation is completely absent for the older stars.  The evolution
of the observed $V_{\phi}$ \vs [Fe/H] trend in the GCS sample
therefore matches our expectations based on the simulation and
strongly suggests that the stellar populations in the solar
neighborhood have been influenced by radial mixing.

\section{Observational Decomposition}
\label{s:obs_decomp}

We now turn our attention to assigning thin or thick disk
membership in a manner analogous to observational studies, i.e. 
either based on kinematics or metallicity. 
We find that classifying stars as members of the thin or thick disk by 
either velocity or metallicity leads to an apparent separation in the 
other property, as observed.  

\subsection{Membership Based on Kinematic Criteria}

One of the observational differences between {\it kinematically
selected} thin and thick disk stars is that the latter have higher
abundance of $\alpha$ elements at a given [Fe/H] \citep[and references
therein]{Bensby2005, Feltzing2006}. This difference is often
interpreted as evidence for different formation histories.  Our
simulations include a calculation of the oxygen abundance following
the prescription by \citet{Raiteri1999}; given that SNII mostly yield
oxygen \citep{Hoffman1999}, we use oxygen as a proxy for all
$\alpha$ elements.  
Because metal yields are not precisely known, 
the normative offset for [$\alpha$/Fe] within the MW simulation does not 
match the Milky Way.  
However, in terms of chemical evolution, the MW simulation gives a good 
qualitative perspective on distributions in [$\alpha$/Fe] space within 
the solar cylinder. 

\begin{figure}[!h]
\epsscale{1}
    \plotone{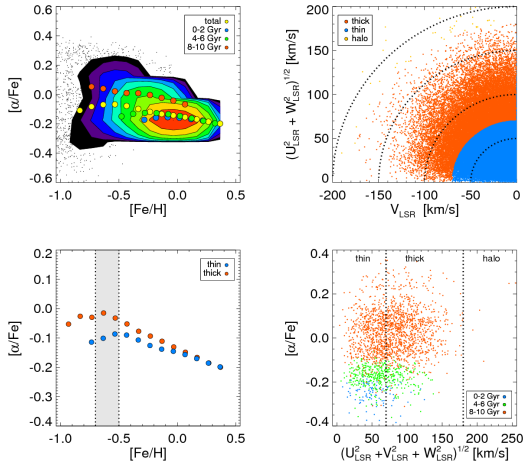}
    \caption{Results of a simple kinematic cut on the local sample: 
             $R = 7 - 9$ $\kpc$, $|z| = 0.0 - 0.3$ $\kpc$.
             Top left: Distribution of [$\alpha$/Fe] for the entire local 
             sample.  The distribution is continuous, not bimodal.  
             The mean [$\alpha$/Fe] for [Fe/H] bins, shown in yellow, 
             qualitatively matches observational data \citep{Bensby2005}.  
             When the sample is decomposed by age, 
             the weighted mean value of old stars is clearly $\alpha$-enhanced               relative to the younger populations.  
             Top right: Toomre diagram.  Stars with $V_{LSR} \ge -70$ km/s
	     are assigned thin disk membership, while stars 
             with $-150 \le V_{LSR} < -70$ km/s are considered thick disk.  
             All other stars are assigned halo membership in agreement with 
             \citet{Nissen2009}.
             Bottom left: the resulting weighted mean distributions for 
             the thin and thick disk populations.  
              The thick disk is $\alpha$-enhanced relative to the 
             thin disk at low [Fe/H].
             Bottom right: stars from just the shaded 
             [Fe/H] cut in the bottom left plot.  
             Stars falling within the ``thick'' disk zone 
             have a higher fraction of old stars relative 
             to the overall population.  
             These old stars are $\alpha$-enhanced.
	     Thus kinematically dividing the stars locally 
	     biases the sample to an older, $\alpha$-enhanced population.}
    \label{f:toomre}
\end{figure}

The top left panel of Figure~\ref{f:toomre} shows the overall
dependence of [$\alpha$/Fe] on [Fe/H] for $R = 7 - 9$ $\kpc$, $|z| =
0.0 - 0.3$ $\kpc$ (\ie the ``solar neighborhood'').  A similar
behavior is seen for the stars at $|z| \sim1$$\kpc$ in agreement with
\citet{Bensby2005}.  When the same sample is separated by age,
distinct portions of the parameter space are covered.  In particular,
at low [Fe/H], old stars show an enhancement of [$\alpha$/Fe] relative
to young stars.

Locally, observed thick disk stars are selected kinematically, rather
than by age \citep{Prochaska2000, Reddy2003, Reddy2006, Allende2004}.
We reproduce the qualitative behavior of observations by following
similar steps with the MW simulation.  The top right panel in
Figure~\ref{f:toomre} shows a Toomre diagram for simulated stars, with
selection cuts motived by \citet{Nissen2009}.  The bottom left panel
shows that these kinematically-selected thin and thick disk
stars from the solar neighborhood show similar bifurcation of
[$\alpha$/Fe] \vs [Fe/H] behavior as young and old subsamples
shown in top left panel. 
In the bottom right panel, stars from just the shaded
[Fe/H] cut in the bottom left plot are examined.
 
Stars falling within the kinematically selected thick disk region 
have a higher fraction of old stars relative to the overall population. 
These old stars are $\alpha$-enhanced as they formed in the interior 
of the disk and radially migrated to their present location. 
Thus kinematically dividing the stars locally biases the sample to an older, 
$\alpha$-enhanced population.
Although the simulation and the data are not in detailed
quantitative agreement, these qualitative results imply that the
differences in [$\alpha$/Fe] \vs [Fe/H] for kinematically selected
thin and thick disk stars may be another consequence of mixing effects
that result from radial migration.

\subsection{Membership Based on [$\alpha$/Fe] Criteria}

Although not as commonly done, it is equally plausible to take the
converse approach; we can assign membership to the thin and thick disk
based on an [$\alpha$/Fe] cut and then study the kinematic and [Fe/H]
distributions that result \citep{Navarro2010, Lee2010}.  
Here we follow the technique outlined by
\citet{Lee2010}, who chemically divided the Galactic disks using SDSS
SEGUE-1 data; we select star particles within $R = 7 - 11$ $\kpc$,
$|z| = 0.3 - 2.0$ $\kpc$ and split them so that stars with
[$\alpha$/Fe] $\geq -0.1$ are assigned thick disk membership.

\begin{figure}[!h]
\epsscale{1}
    \plotone{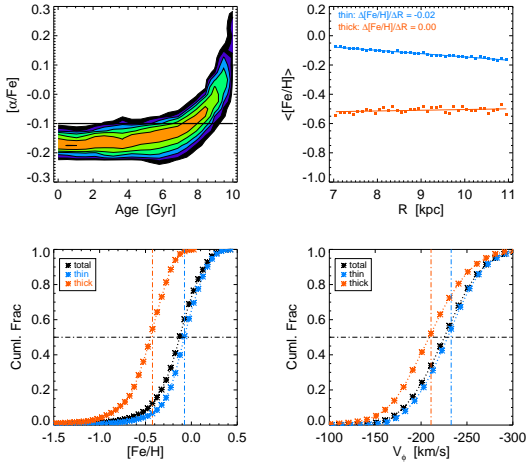}
    \caption{Thin and thick disk membership assigned based 
              on [$\alpha$/Fe]; the sample includes star particles 
              within $R = 7 - 11$ $\kpc$, $|z| = 0.3 - 2.0$ $\kpc$.
              Top left: mass weighted contour plot of 
              [$\alpha$/Fe] $\vs$ Age in logrithmically spaced bins 
              (low to medium to high: black to green to orange).
              Overplotted for reference is the dividing line 
              [$\alpha$/Fe] $= -0.1$; stars with [$\alpha$/Fe] 
              $\ge -0.1$ are considered $\alpha$-enhanced and are assigned 
              thick disk membership. Clearly the majority of these stars 
              are quite old.  Stars with [$\alpha$/Fe] $< -0.1$ are 
              considered thin disk members and sample a wide range of ages.
	      Counterclockwise from top right to bottom left: resulting thin 
              and thick disk trends.
              Notably, the thick disk stars are metal poor with no gradient 
              in R and lag the rotation of the thin disk stars.  
              Despite the appearance of these trends here, 
              there is no distinct thick disk population in 
              the MW simulation.}
    \label{f:connie}
\end{figure}

Assigning membership based on 
[$\alpha$/Fe] effectively divides the disk into two populations: old stars 
and young -- intermediate age stars, as can be seen clearly in the
top left panel of Figure~\ref{f:connie}.  
Why is this the case?  Star particles are born
$\alpha$-enhanced if they form in a region with a high local SFR and 
little SNIa pollution.  
Since most old stars originated near the center of the disk, and
that region is where the SFR was high, these stars are naturally
$\alpha$-enhanced.

We compare our results to three trends discussed in \citealt{Lee2010}:
the radial metallicity gradient, distribution in $V_{\phi}$ and
distribution in [Fe/H], which we show clockwise from top right in
Figure~\ref{f:connie}.  The top right panel illustrates the best fit
radial metallicity gradients as derived from the mass weighted mean
value of [Fe/H].  Cutting by [$\alpha$/Fe] results in no trend
(slope $\sim$$0\kpcdex$) in the thick disk and a negative
trend in the thin disk (slope $\sim$$-0.2\kpcdex$); this is similar to
the thick disk slope $=0\kpcdex$ and thin
disk slope $=-0.3\kpcdex$ observed by \citet{Lee2010}.  The bottom right
panel illustrates the cumulative $V_{\phi}$ distributions that result
for the MW simulation when stars are separated by [$\alpha$/Fe]: the
thin and thick disk trends are offset by $\sim$$25\kms$, which is
qualitatively similar to, if quantitatively smaller than, the
$\sim$$50\kms$ offset found by \citet{Lee2010}.  Finally the bottom
left panel shows the cumulative [Fe/H] distributions; we find the thin
and thick disk trends offset by $\sim$$0.35$ dex, not dissimilar from
the observed $\sim$$0.4$ dex offset \citep{Lee2010}.

Therefore, by adopting a kinematic or [$\alpha$/Fe] selection criteria 
used by observers, we are able to reproduce an apparent separation in the 
other property, despite the fact that there are no distinct populations
in the model galaxy.

\section{Comparison with Previous Theoretical Work}
\label{s:schoenrich}

Recently, \citet{Schoenrich2009} investigated how radial migration and
chemical evolution shape the solar neighborhood, by incorporating for
the first time a prescription for radial migration in a semi-analytic
model of Galactic chemical evolution. Their model represented a disk
in which star formation commenced at all radii simultaneously (\ie
without inside-out growth), with radially varying star formation rates
set to yield a disk with an appropriate scale length. The guiding
center radii of the stars in their model changed (i.e. stars migrated
radially) according to a parametrized probabilistic prescription whose
normalization was left as a free parameter in the model. The vertical
structure of their disk was determined based on the assumption that
coeval stars comprise an isothermal population with a velocity
dispersion given by local observational constraints.  Because stars
migrating from the inner disk retain their velocity dispersions but
encounter a lower restoring potential in the outer disk, they populate
the disk away from the plane.  Hence, \citet{Schoenrich2009} showed
that a thickened component may result simply by (the inevitable)
radial migration, and by fitting the model they also reproduced many
of the canonical features of the thick disk (\ie enhanced
[$\alpha$/Fe] ratios, older ages, lower metallicites, rotational lag), 
similar to what we have shown in the previous Section.

\begin{figure}[!h]
\epsscale{1}
    \plottwo{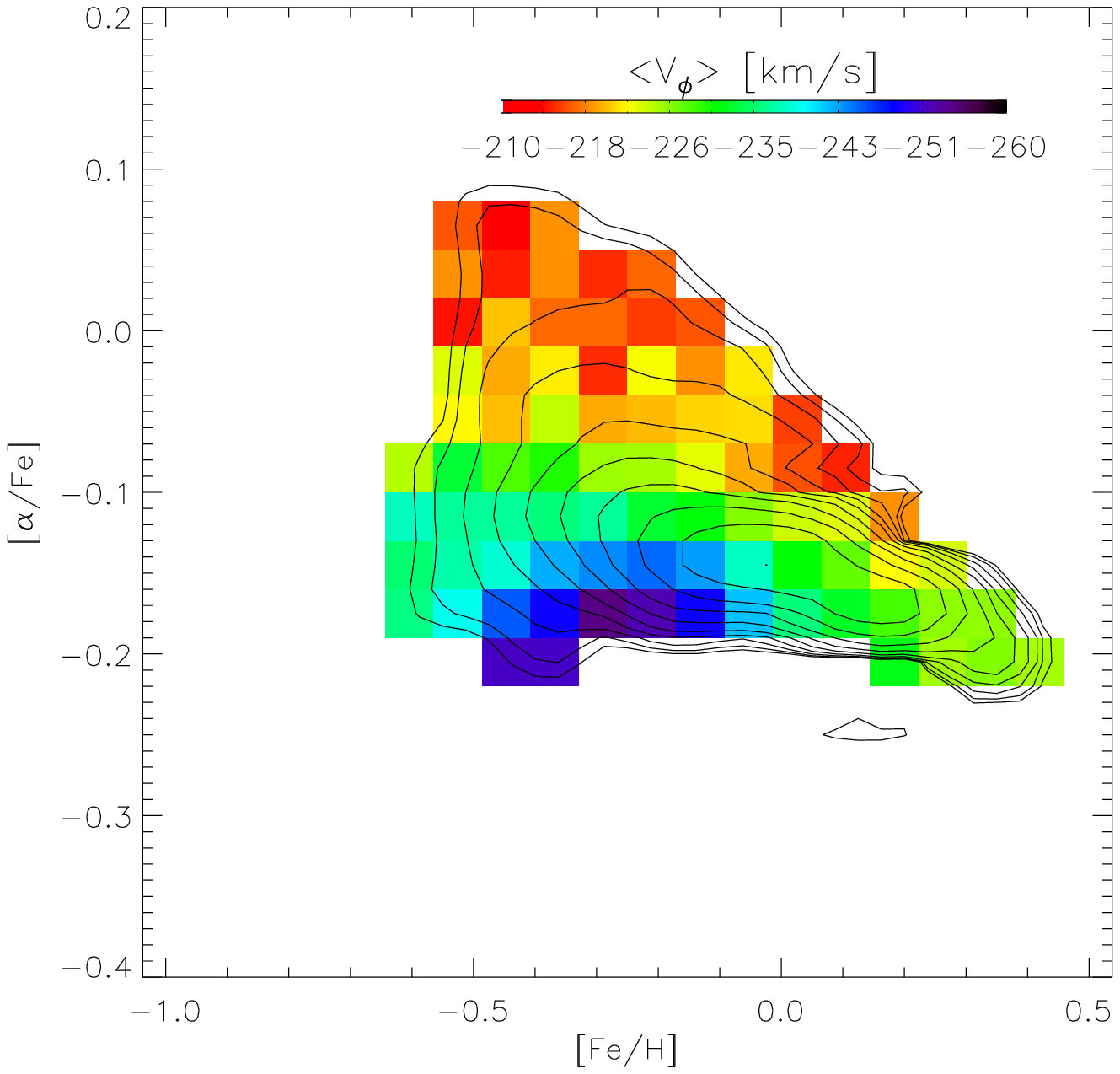}
            {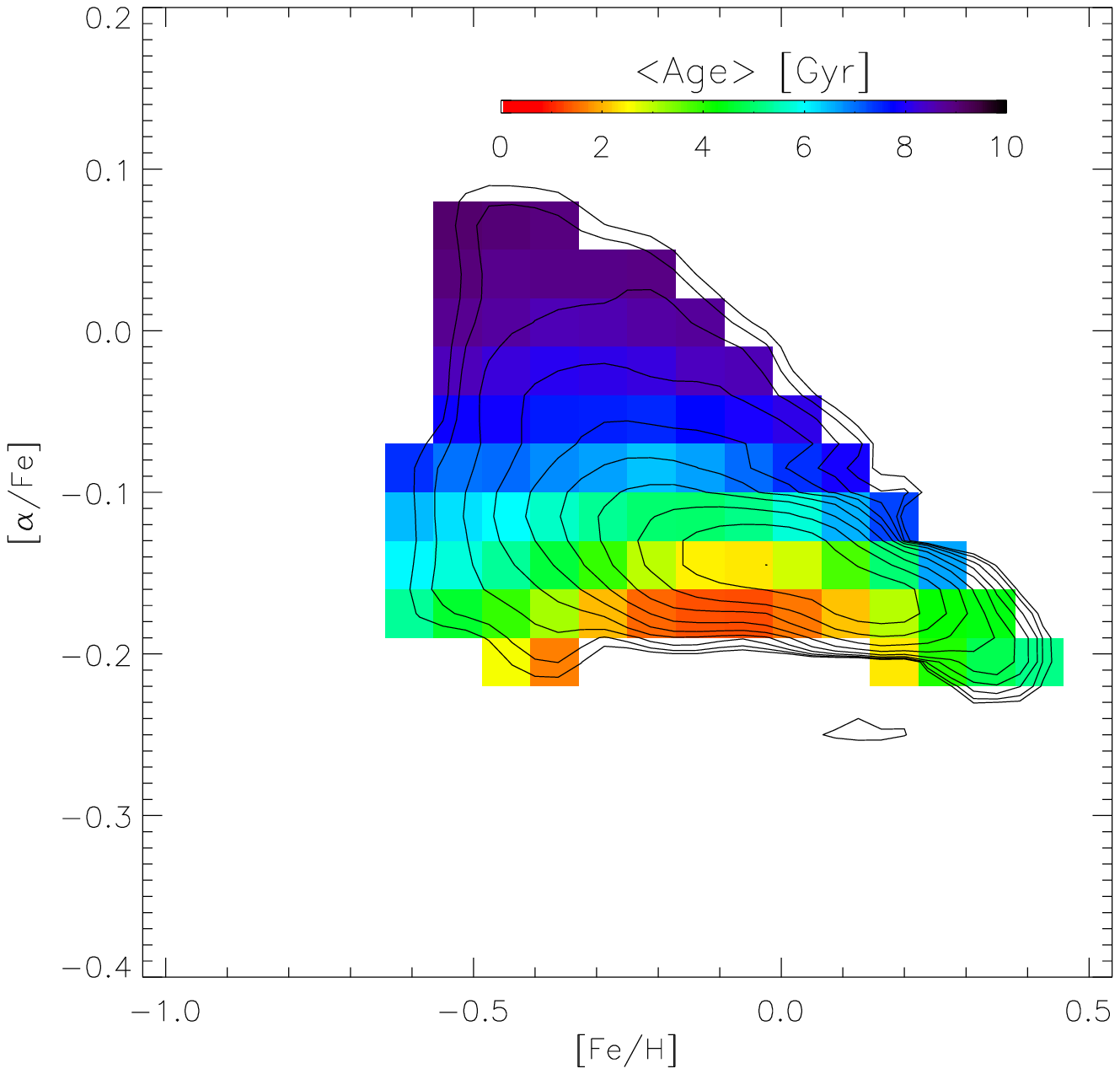}
    \plottwo{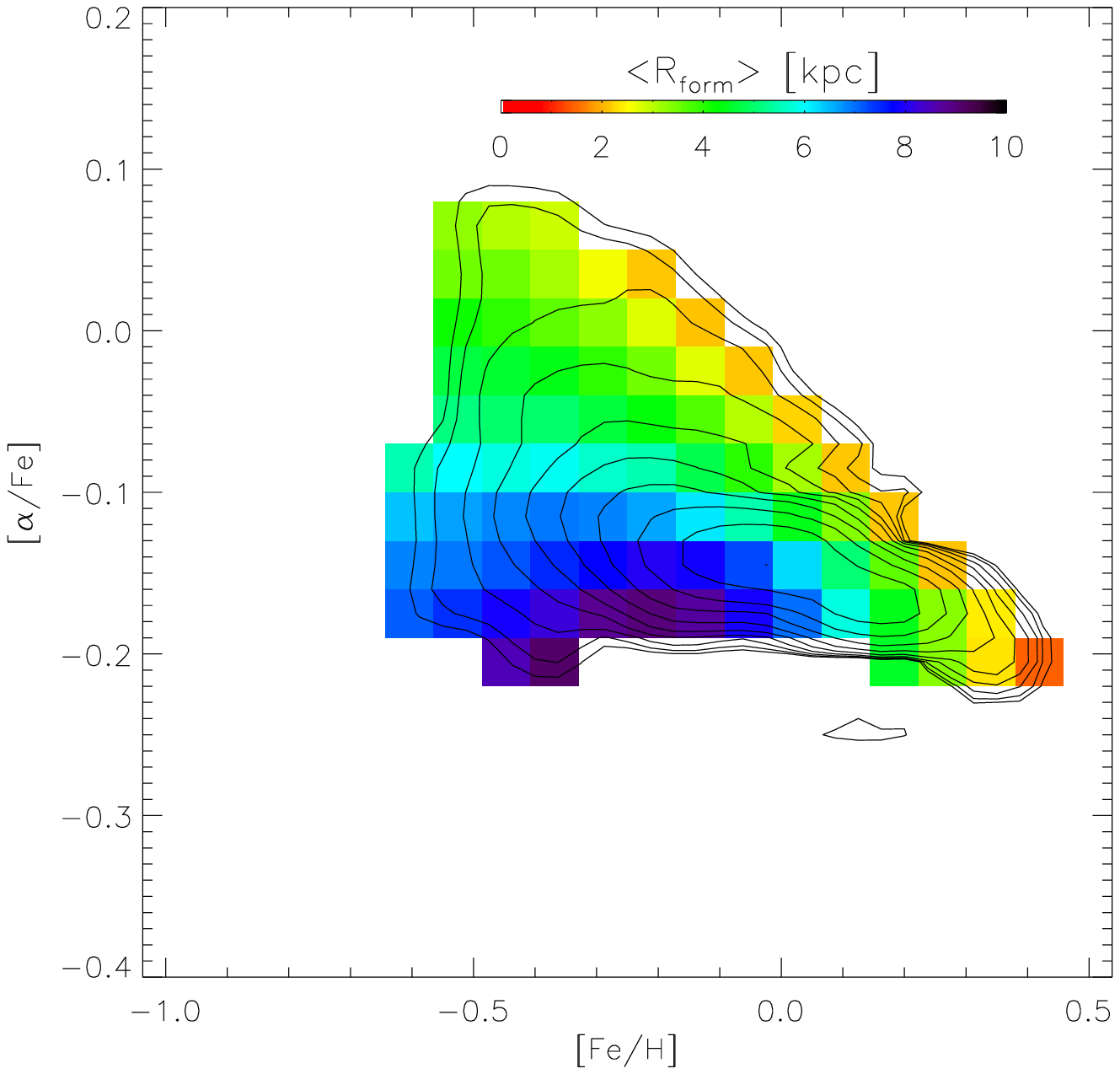}
            {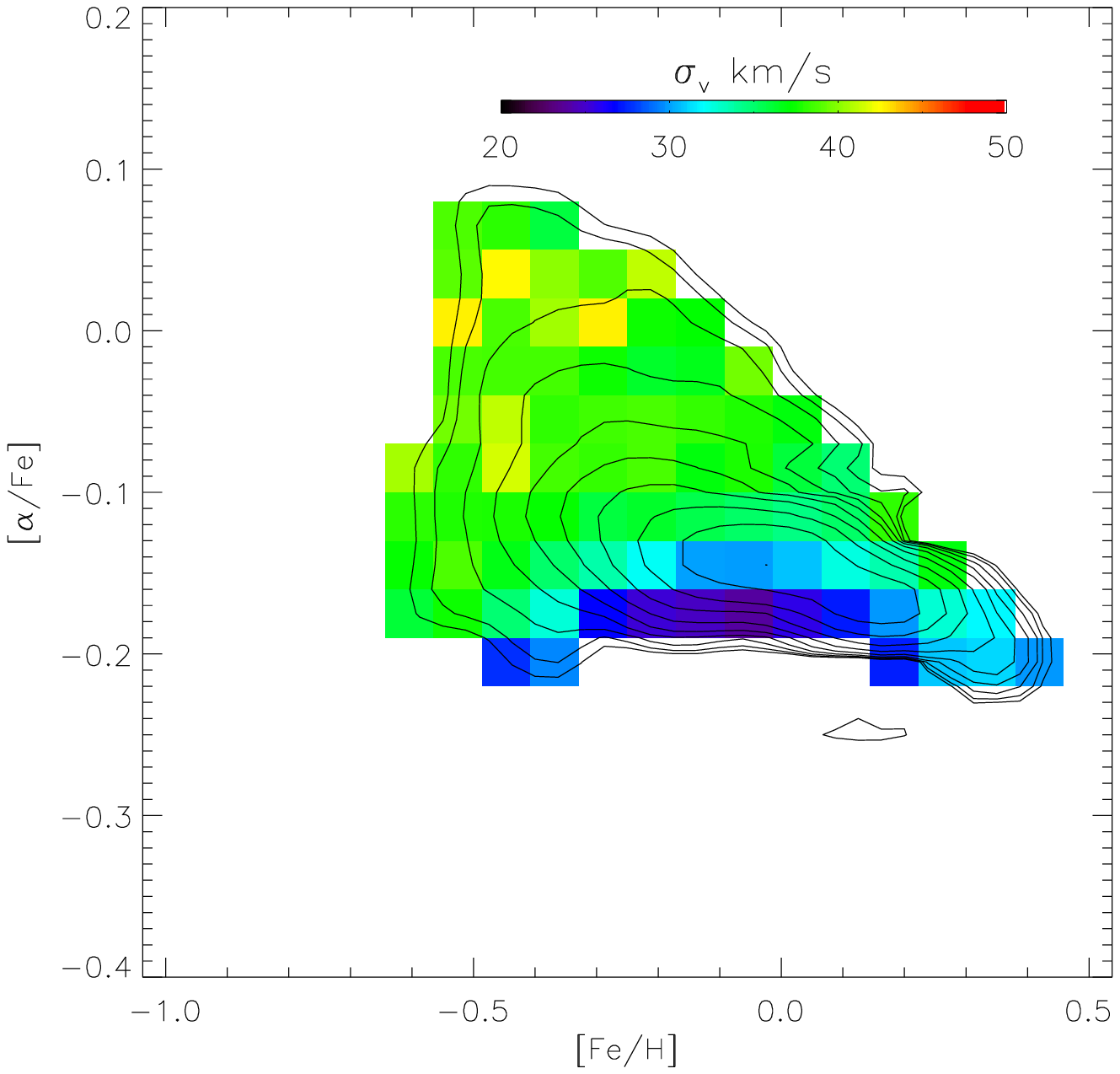}
    \caption{The panels clockwise from top left show distributions 
             of mean $V_{\phi}$, age, $\sigma_{V}$, and $R_{form}$ in the 
             [$\alpha$/Fe] \vs [Fe/H] plane for particles in the midplane 
             ($|z| < 0.3 \kpc$) at the solar radius ( $7 < R $ [kpc] $< 9$).  
             The black contours are logarithmically spaced and indicate 
             particle density, while the colors correspond to the mean of the 
             specified quantity.  Only cells containing at least 200 particles 
             are shown.}
    \label{f:rok_plot}
\end{figure}

It is therefore reassuring that our results presented here agree
qualitatively with those of \citet{Schoenrich2009}, given that our
modeling methods are entirely different, and that our model was not specifically
tuned to the Galaxy.  However, we also find subtle yet crucial
differences.  Figure 4 of \citet{Schoenrich2009} shows that due to
radial mixing, a population that shares the same kinematics can show
immense variations in its chemical composition. The top left panel of
Figure~\ref{f:rok_plot} shows that our model also yields stellar
populations with the same kinematics but very different chemistry, 
consistent with recent observational work \citep{Navarro2010}.
However, we note the contours of mean $V_{\phi}$ in this plane are almost
orthogonal in our simulation compared to those shown in Fig. 4 of
\citet{Schoenrich2009}, even though the age structure is very similar;
at fixed [Fe/H] age increases monotonically with rising [$\alpha$/Fe]
(middle panel of Figure~\ref{f:rok_plot} and Figure 5 of
\citealt{Schoenrich2009}).

A hint of a reason for this discrepancy is provided by scrutinizing
the metal-rich end of these figures. In our model, the
[$\alpha$/Fe]-deficient metal-rich population originated in the interior of
the disk (see the rightmost panel of Figure~\ref{f:rok_plot}) and has
migrated to the present radius without very much heating.  This is
apparent from the fact that the mean $V_{\phi}$ for this population is
only very slightly lagging the LSR ($\sim$$-240~\kms$).  On the other
hand, in the model of \citet{Schoenrich2009}, that same population shows
considerable lag from the LSR. This discrepancy implies that there are
qualitative differences in the treatment of radial mixing between
their prescription and our simulation. 

On the other hand, the differences in the velocity structure in this
plane must also be a result of their assumption that the entire disk
begins forming stars at once with a peak in star formation occurring
$\sim$10 Gyr ago everywhere. In our model, it is impossible to have an
old metal poor star which formed at the solar radius - \textit{all}
of these stars must have migrated to their present position because the solar
neighborhood in our simulation does not exist $\sim$$9$~Gyr ago. 

\citet{Haywood2008} argued that the existing solar neighborhood
samples (\eg the GCS) show signatures of radial
mixing as proposed by \citet{Sellwood2002}.  In our sample shown in
Figure~\ref{f:rok_plot}, the metal-poor, low [$\alpha$/Fe] stars have high
velocities and young ages, and are at or near peri-galactic passage from the
outer disk into the solar neighborhood sample.  This is consistent
with the arguments put forth by \citet{Haywood2008} as observational
evidence of radial mixing.  However, we point out that the large 
velocities of this tail in the distribution signify that their orbits
have merely been heated, and these stars have not migrated via the
corotation scattering mechanism over any significant distance.
Instead, it is the presence of significantly metal-enriched stars on
kinematically inconspicuous orbits (\ie kinematically cool) that
should be considered as clear evidence of migration.

\section{Predictions for upcoming surveys}
\label{s:predict}

We have demonstrated that we can understand the lack of a correlation
between $V_{\phi}$ and [Fe/H] as a consequence of radial migration
effects; the significance of these effects became clear when we
decomposed the MW simulation within the solar cylinder by age.
Unfortunately age is not easily accessible observationally.  However,
$\alpha$ is a directly measurable quantity and, as we have shown, for
the oldest stars, $\alpha$-enhancement is a reasonable proxy for age
(see top left panel of Figure~\ref{f:connie}).  Thus we now reassess
the relationship between $V_{\phi}$ and [Fe/H] by decomposing
Figure~\ref{f:vphi_feh_decomp} in cuts of [$\alpha$/Fe] so that we can
make a testable prediction for upcoming observational surveys.

\begin{figure}[!h]
\epsscale{1}
    \plotone{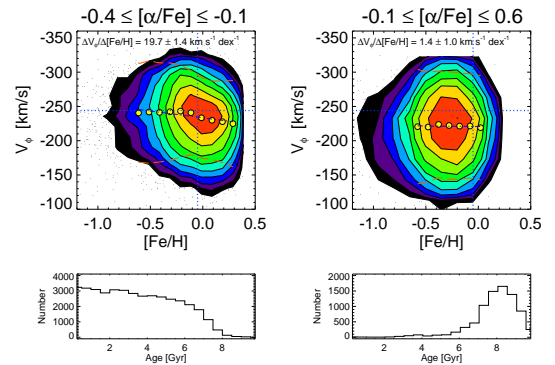}
    \caption{$V_{\phi}$ \vs [Fe/H] for two broad bins of [$\alpha$/Fe] for 
             the volume spanning $|z| = 0.5 - 1.0$ $\kpc$, $R = 7 - 9$ $\kpc$. 
             Top left panel:
             $-0.4 \leq [\alpha$/Fe]$\leq -0.1$.  Note the knee in the 
             distribution at $\sim$$-0.2$ dex. 
             Bottom left panel: the distribution of ages sampled by
             this [$\alpha$/Fe] cut. 
             Top and bottom right panel: analogous to the left panel but for 
             $-0.1 \leq [\alpha$/Fe]$\leq 0.6$.  
             Note the histogram is dominated by stars older than 7 Gyr old 
             and there is no trend in the $V_{\phi}$ \vs [Fe/H] figure.  
             This is a clear indication of the importance of radial migration 
             in this volume.}
    \label{f:local_alpha_cuts}
\end{figure}

Figure~\ref{f:local_alpha_cuts} shows $V_{\phi}$ $\vs$[Fe/H] for the
``transition zone'' considered in Section~\ref{s:solar} ($|z| = 0.5 -
1.0$ $\kpc$, $R = 7 - 9$ $\kpc$), as split by two broad bins in
[$\alpha$/Fe].  Below each $V_{\phi}$ $\vs$[Fe/H] plot is a histogram
of the ages represented in the given bin.  In the right panel, the
high [$\alpha$/Fe] sample contains almost exclusively old stars
($84\%$ older than 7 Gyr); the corresponding $V_{\phi}$ $\vs$[Fe/H]
figure shows no trend between the quantities (best fit slope
$\sim$$1\kmsdex$).  In contrast to this, the left panel, the low
[$\alpha$/Fe] content bin has few old stars ($< 5\%$ older than 7 Gyr)
while sampling young to intermediate aged stars relatively equally
($31\%$ between 0 and 2 Gyr old, and $25\%$ between 4 and 6 Gyr old).
Here there is a trend between $V_{\phi}$ and [Fe/H], with an overall
best fit value of $\sim$$20\kmsdex$.  We note that this [$\alpha$/Fe] 
decompositon works equally well in the solar neighborhood as the 
``transition zone''.

For the low [$\alpha$/Fe] cut, we have fit the mass weighted mean
values with a single linear fit; however it is equally plausible to
fit two lines here: $-0.6$ dex [$\leq \alpha$/Fe] $\leq -0.2$ dex and
the other $-0.2$ dex [$\leq \alpha$/Fe] $\leq 0.3$ dex.  In that case,
the linear fit to low [Fe/H] shows no trend while the linear fit to
high [Fe/H] shows a strong trend.  Notably, the portion spanning low
[Fe/H] is dominated by the intermediate aged stars while the portion
spanning high [Fe/H] is dominated by a younger population.  This
``knee'' is a persistent feature; when the upper limit on this
[$\alpha$/Fe] cut is lowered, fewer intermediate age stars are
sampled, and the knee in the trend shifts to lower [Fe/H].  Note, the
trend is evident here because on average the stars in this [Fe/H]
space have experienced less radial mixing than older stars within the
same spatial volume.  Thus, it is possible to recover this signature
of radial mixing even in the absence of age estimates, but with
knowledge of [$\alpha$/Fe], [Fe/H], and $V_{\phi}$ for an unbiased
population of stars located out of the midplane.

To date, $\alpha$ measurements have only been accessible to small 
targeted samples; our initial comparison with a compilation of all 
currently available data is particularly encouraging \citep{Navarro2010}.   
We note current work, like that 
of the SDSS SEGUE collaboration, aims to obtain
a large, well-calibrated [$\alpha$/Fe] dataset.
We eagerly anticipate the application of cuts on [$\alpha$/Fe] to an
unbiased population of stars that fall within the region considered by
I08 or indeed any large sample within the solar cylinder; such an analysis 
would further elucidate whether radial mixing has played an important role in shaping
the distribution of the Milky Way stars over time.

\section{Conclusions}
\label{s:concl}

We have used an N-body model designed to mimic the quiescent formation
and evolution of a Milky Way--type galactic disk (Ro\v{s}kar et
al. 2008ab) to interpret recent SDSS-based observational constraints
on the structure of the Milky Way disk. Traditional simplistic
decomposition of thin and thick disks as two distinct populations of
metallicity and rotational velocity predicts a strong correlation of
these quantities at $\sim$1 kpc from the Galactic plane. Thanks to
nearly complete flux-limited SDSS stellar samples with both rotational
and metallicity measurements, the expected correlation was strongly
ruled out (Ivezi\'{c} et al. 2008).  On the other hand, the MW simulation we presented here, while
not a detailed model of the Milky Way, produces good qualitative
agreement with the data and sheds new light on the origin and
evolution of the observed disk structure.  Of particular importance is
the role of radial migration in mixing stars born
throughout the disk into the solar neighborhood.

The properties of the Ro\v{s}kar et al. model for the overall spatial,
metallicity and kinematic distributions of the Milky Way stars are in
qualitative agreement with the SDSS data (Juri\'{c} et al. 2008;
Ivezi\'{c} et al. 2008; Bond et al. 2010).  While there are
quantitative differences in the spatial gradients of these
distributions, as well as in their detailed behavior, even this
qualitative agreement is remarkable because the simulation was not
fine tuned to match the Milky Way. Not only does the model reproduce
the observed change of slope in the counts of disk stars as a function
of distance from the Galactic plane (the original motivation for
introducing a separate thick disk component), but it also predicts the
gradients in metallicity and rotational velocity. Furthermore, the
metallicity and rotational velocity are uncorrelated in this model for
appropriately volume-selected subsamples of stars, in agreement with
observations.

The robust qualitative agreements between the data and model
predictions motivate the use of model quantities inaccessible to
observations, such as the stellar age and the ISM metallicity at the
time and position of stellar birth, to interpret recent SDSS results.
In particular, the lack of correlation between the metallicity and
rotational velocity at $\sim$1 kpc from the Galactic plane can be
understood as due to complex interplay between the ISM metallicity at
the time and position of stellar birth, and the subsequent secular
evolution largely driven by spiral arms.  

No {\it a priori} assumptions about the disk's structure are
incorporated in the model -- yet it reproduces the main observational
results which motivated decomposition of the disk into two presumably
distinct components. The absence of mergers in this model implies that
they are not required to explain the overall disk structure.  While
merger remnants are detected within the Milky Way disk (e.g. J08),
their influence is apparently well localized.  Instead, the Milky Way
disk can be viewed as a single complex structure with an age
distribution that is a strong function of position in the Galaxy due
to radial migration of stars. Thus even if a primordial thick disk is
present, having formed via accretion/external heating, it is likely to
be substantially polluted by migrating disk stars.

The same mixing effects are likely responsible
for the observed differences in $\alpha$ element abundance between
{\it kinematically selected} thin and thick disk stars. By adopting
kinematic selection criteria used by observers, we are able to
reproduce distinctive [$\alpha$/Fe] \vs [Fe/H] trends similar to
those seen in the data, despite the fact that there are no distinct
populations in the model galaxy.  

We look forward to the improved data derived from the emerging
generation of surveys such as SEGUE \citep{Rockosi2009} and 
APOGEE \citep{Majewski2010}.  
Key to garnering a deeper understanding of the importance of 
radial migration in the Milky Way evolution is gathering both 
precise age determinations and detailed chemical compositions.  
We are optimistic that this study will lead to further observational 
and theoretical work.

\section{Acknowledgments}
\label{s:ack}
We thank our numerous SDSS (www.sdss.org) collaborators for their 
valuable contributions and helpful discussions.
S. Loebman and \v{Z.} Ivezi\'{c} acknowledge support by NSF grants
AST-0707901 and AST-1008784 to the University of Washington, and by
NSF grant AST-0551161 to LSST for design and development
activity. \v{Z.} Ivezi\'{c} acknowledges support by the Croatian
National Science Foundation grant O-1548-2009.
This research was supported in part by the NSF through TeraGrid 
resources provided by TACC and PSC. 
R. Ro\v{s}kar and T. R. Quinn were supported by the NSF ITR grant PHY-0205413 
at the University of Washington.


\appendix

\section{Conflicting Observational Claims}
\label{s:spagna}

Recently, \citet{Spagna2010} found that the rotational velocity for
disk stars is correlated with metallicity for $-1$ $<$ [Fe/H] $<$
$-0.5$ at $1$ $\kpc$ $<$ $z$ $<$ $3$ $\kpc$.  Notably, they found a
gradient within $40$ -- $50$ $\kmsdex$, such that more metal-poor
stars rotate more slowly.  Their claim is in direct conflict with
several other observational studies \citep[][I08, B10]{Carollo2010}.

\begin{figure}[!h]
  \epsscale{1}
  \plottwo{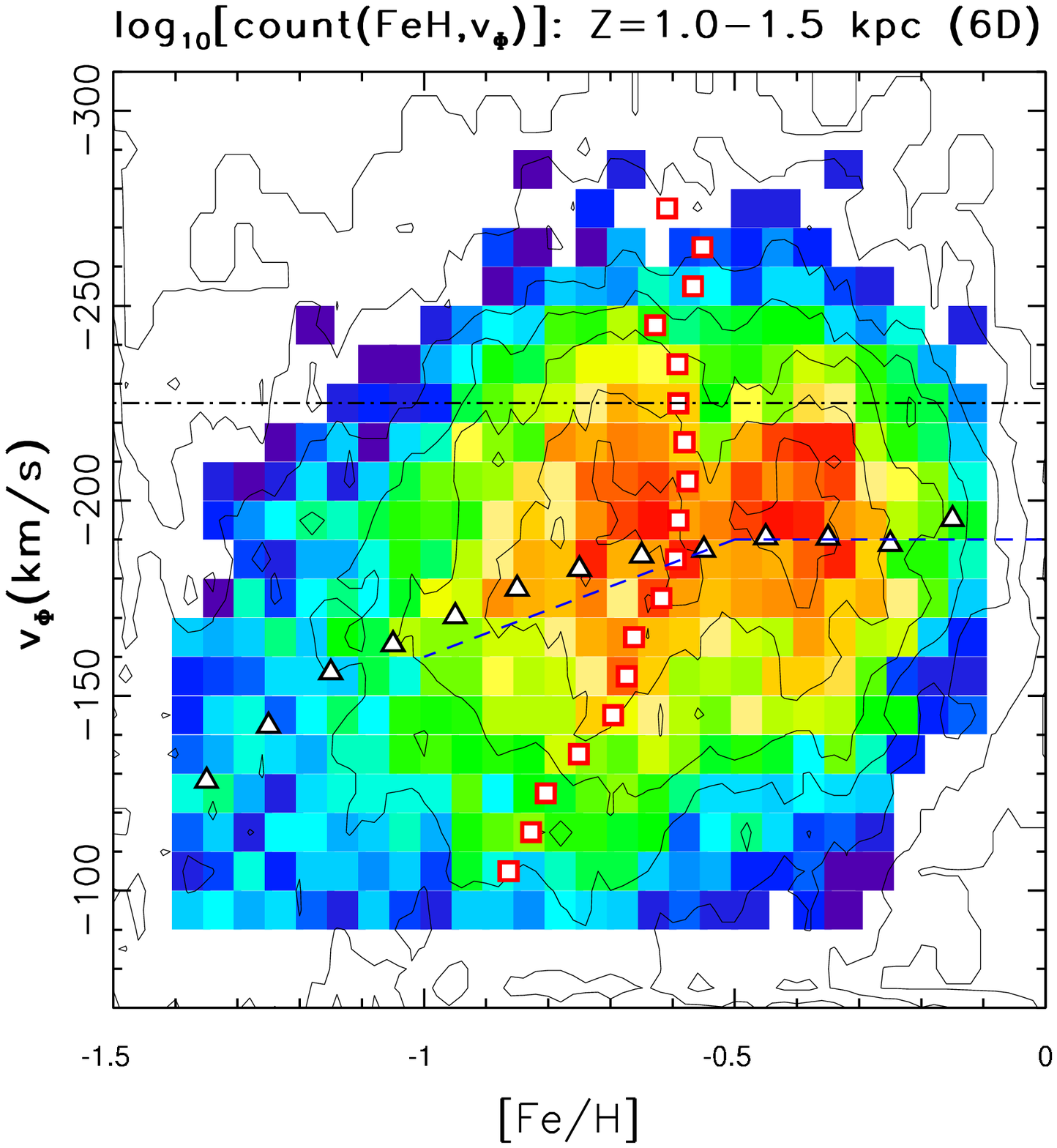}{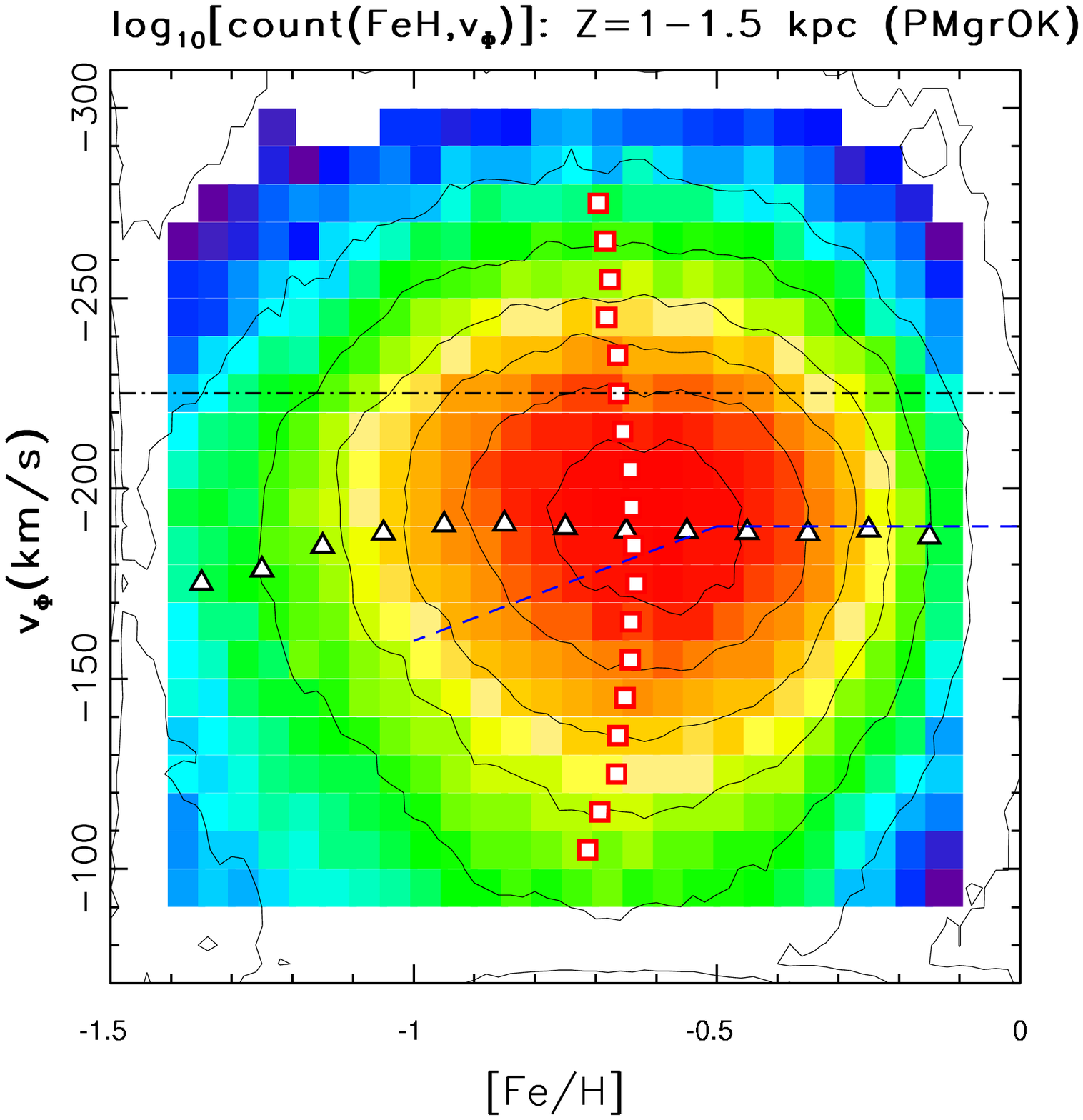}
  \caption{Left panel: 
           $\sim 16,000$ stars selected from SDSS DR7 spectroscopic sample 
           with $g-r = 0.2 - 0.6$ from $z = 1.0 - 1.5$ $\kpc$.
           The stellar number density is shown as the color-coded map 
           (low to high: blue to red) and by the contours. 
           Triangles are the median $V_{\phi}$ for bins of [Fe/H], and
           squares are the median [Fe/H] for bins of $V_{\phi}$.
           \citet{Spagna2010} results are overplotted 
           with a dashed blue line for reference.
           Right panel:
           Analogous to plot in right panel 
           but for full photometric sample: 
           $\sim 124,000$ stars with $g-r = 0.2 - 0.6$ 
           and $z = 1.0 - 1.5$ $\kpc$, selected from the
	   meridional plane defined by $l \sim 0\,^{\circ}$ or 
           $l \sim 180\,^{\circ}$ (See Section 3.2 in Bond et al. 2010).}
  \label{f:spagna}
\end{figure}

We reconsider the \citet{Spagna2010} findings using SDSS DR7 data  
to try to understand the differences between these observational 
findings.
We note a direct comparison between I08 and \citet{Spagna2010} is non-trivial 
as \citet{Spagna2010} use their own proper motion measurements based on the 
GSC-II catalog and do not provide a comparison to the 
\citet{Munn2004} proper motions on a star-by-star basis.
Additionally, their color selection is more generous ($0.0<g-r<0.9$) 
than previous studies ($0.2<g-r<0.6$, I08). 
This more generous selection lets in BHB stars, as well as red stars 
where [Fe/H] reliability decreases.
Despite differences in selection criteria, we can reproduce \citet{Spagna2010} 
results when using the SDSS \textit{spectroscopic} sample 
(see left panel, Figure \ref{f:spagna}).

While the left panel of Figure \ref{f:spagna} 
only considers a narrow bin of  
$z = 1.0 - 1.5$ $\kpc$, we find the best fit lines 
are reproducible for other bins as well.
Notably, the median $V_{\phi}$ as a function of [Fe/H] closely 
follows the \citet{Spagna2010} 
result. 
We also reproduce a bimodal distribution of stars in the metallicity 
direction, with modes at [Fe/H]$ \sim -0.65$ and [Fe/H] $\sim -0.4$ 
\citep[see Section 3.2,][]{Spagna2010}.

However, we detect no $V_{\phi}$ --- [Fe/H] correlation when we
consider a complete sample selected in the meridional plane
($l\sim0\,^{\circ}$ or $l\sim180\,^{\circ}$), where proper motion
alone suffices to measure rotational velocity (see right panel, 
Figure~\ref{f:spagna}).  This sample is essentially complete in
selected color-distance limits and thus not subject to strong
selection effects present in the SDSS spectroscopic sample.  This
sample shows negligible dependence of the median $V_{\phi}$ on
[Fe/H] for [Fe/H]$ > -1.0$ (one can also see the bias due to halo
stars for [Fe/H]$ < -1.0$).

It is highly likely that the \citet{Spagna2010} results are caused by
selection biases in the SDSS spectroscopic sample.  Notably, when we
fit two gaussians (one for the disk and halo) to the $V_{\phi}$ and
[Fe/H] distributions, about $45\%$ of the spectroscopic sample are
halo stars (consistent with \citet{Spagna2010} within $\sim 5\%$
errors), while halo stars make up only $8\%$ of the complete sample.
And when we compare the two panels in Figure \ref{f:spagna} 
we are using exactly the same volume and exactly
the same measurements: the only difference between the samples is that
the spectroscopic sample includes only $\sim 2\%$ of all the stars,
with the selection probability about 10 times higher for halo stars
than for disk stars.

\end{document}